\documentclass{article} 
\usepackage{lipsum}
\usepackage{multirow}
\usepackage{booktabs}
\usepackage{multirow}
\usepackage{multicol}
\usepackage{adjustbox}
\usepackage{tcolorbox}
\usepackage{tabularx}
\usepackage{tikz}
\usepackage{graphicx}
\usepackage[FIGTOPCAP]{subfigure}

\usepackage{amsmath,amsfonts,bm}

\def\eqref#1{equation~\ref{#1}}

\def\1{\bm{1}}

\DeclareMathAlphabet{\mathsfit}{\encodingdefault}{\sfdefault}{m}{sl}
\SetMathAlphabet{\mathsfit}{bold}{\encodingdefault}{\sfdefault}{bx}{n}

\usepackage{hyperref}
\usepackage{url}
\usepackage{lineno}
\usepackage{array, makecell}
\usepackage{algorithm}
\usepackage{algpseudocode}
\usepackage[numbers]{natbib}
\usepackage{geometry}
\usepackage{amssymb}
\usepackage{chngcntr}
\usepackage{subcaption}
\usepackage{enumitem}
\usepackage{overpic}

\usepackage{float} 
\title{\textsc{LLMatDesign: Autonomous Materials Discovery with Large Language Models}}

\author{Shuyi Jia$^\dag$, Chao Zhang$^\dag$, Victor Fung$^{\dag\text{\textsection}}$\\
{\normalsize $^\dag$Computational Science and Engineering, Georgia Institute of Technology, Atlanta, GA, USA} \\
{\normalsize $^{\text{\textsection}}$Corresponding author: \texttt{victorfung@gatech.edu}}
}
\date{}

\newcolumntype{Y}{>{\centering\arraybackslash}X}

\begin{document}

\maketitle

\begin{abstract}
\noindent Discovering new materials can have significant scientific and technological implications but remains a challenging problem today due to the enormity of the chemical space. Recent advances in machine learning have enabled data-driven methods to rapidly screen or generate promising materials, but these methods still depend heavily on very large quantities of training data and often lack the flexibility and chemical understanding often desired in materials discovery. We introduce LLMatDesign, a novel language-based framework for interpretable materials design powered by large language models (LLMs). LLMatDesign utilizes LLM agents to translate human instructions, apply modifications to materials, and evaluate outcomes using provided tools. By incorporating self-reflection on its previous decisions, LLMatDesign adapts rapidly to new tasks and conditions in a zero-shot manner. A systematic evaluation of LLMatDesign on several materials design tasks, \textit{in silico}, validates LLMatDesign's effectiveness in developing new materials with user-defined target properties in the small data regime. Our framework demonstrates the remarkable potential of autonomous LLM-guided materials discovery in the computational setting and towards self-driving laboratories in the future.
\end{abstract}

\section{Introduction}\label{sec:intro}

Discovering novel materials with useful functional properties is a longstanding challenge in materials science due to the vast and diverse composition and structure space these materials can inhabit\citep{davies2016computational, oganov2019structure}. Traditional approaches to materials discovery often involve exhaustively screening materials via lab-based experiments or \textit{in silico} simulations, which can be time-consuming and resource-intensive\citep{liu2017materials, hautier2012computer, pyzer2015high}. Recent advancements have introduced machine learning surrogate models to predict material structures and properties \citep{chen2022universal, merchant2023scaling}, as well as generative modeling techniques to propose novel materials \citep{hoffmann2019data, court20203, xie2021crystal, long2021constrained, ren2022invertible, fung2022atomic, zeni2023mattergen}. However, these data-driven methods rely heavily on extensive training datasets, generally derived from density functional theory (DFT) calculations. These methods are less useful in most instances where such data is unavailable, or when only a limited budget exists to perform experiments or high fidelity simulations. In contrast, a human expert would be far more effective here by being able to draw from domain knowledge and prior experiences, and reason from limited examples. Therefore, a different materials design paradigm is needed in these situations where models should be developed to exhibit similar proficiencies as human experts.

Fueled by ever-expanding textual datasets and significant increases in computing power, large language models (LLMs) have witnessed a meteoric rise in capabilities and usage in recent years. More broadly, the remarkable performance of LLMs across diverse tasks they have not been explicitly trained on has sparked a burgeoning interest in developing and utilizing LLM-based agents capable of reasoning, self-reflection, and decision-making\citep{wei2022chain, huang2022towards, li2022pre}. These autonomous agents are typically augmented with tools or action modules, empowering them to go beyond conventional text processing and directly interact with the physical world, such as robotic manipulation \citep{ahn2022can, huang2023voxposer} and scientific experimentation \citep{boiko2023autonomous, bran2023chemcrow}. As the capabilities of LLMs and LLM-based autonomous agents continue to expand, they are increasingly being recognized for their potential in scientific domains, particularly in chemistry \citep{ai4science2023impact}. This surge in interest stems from the fact that the majority of information in chemistry exists as text, aligning closely with the text-centric nature of LLMs \citep{mirza2024large}. For instance, recent studies have demonstrated the use of LLMs to extract chemical reaction information \citep{fan2024openchemie, ai2024extracting}, predict chemical properties \citep{zhong2024benchmarking, xie2024fine, jablonka2024leveraging, ock2023catalyst}, and generate crystal structures \citep{flam2023language, antunes2023crystal, gruver2024fine}, among many other applications. In particular, chemical research, such as materials discovery, traditionally hinges on human expertise and experience encapsulated in scientific publications. LLMs, capable of ingesting vast quantities of these publications beyond human capacity, have the potential to act as intelligent copilots that might be able to extract key insights, uncover hidden patterns, and propose novel methodologies, thereby accelerating scientific progress \citep{mirza2024large}.

\begin{figure}[h]
    \centering
    \includegraphics[width=\textwidth]{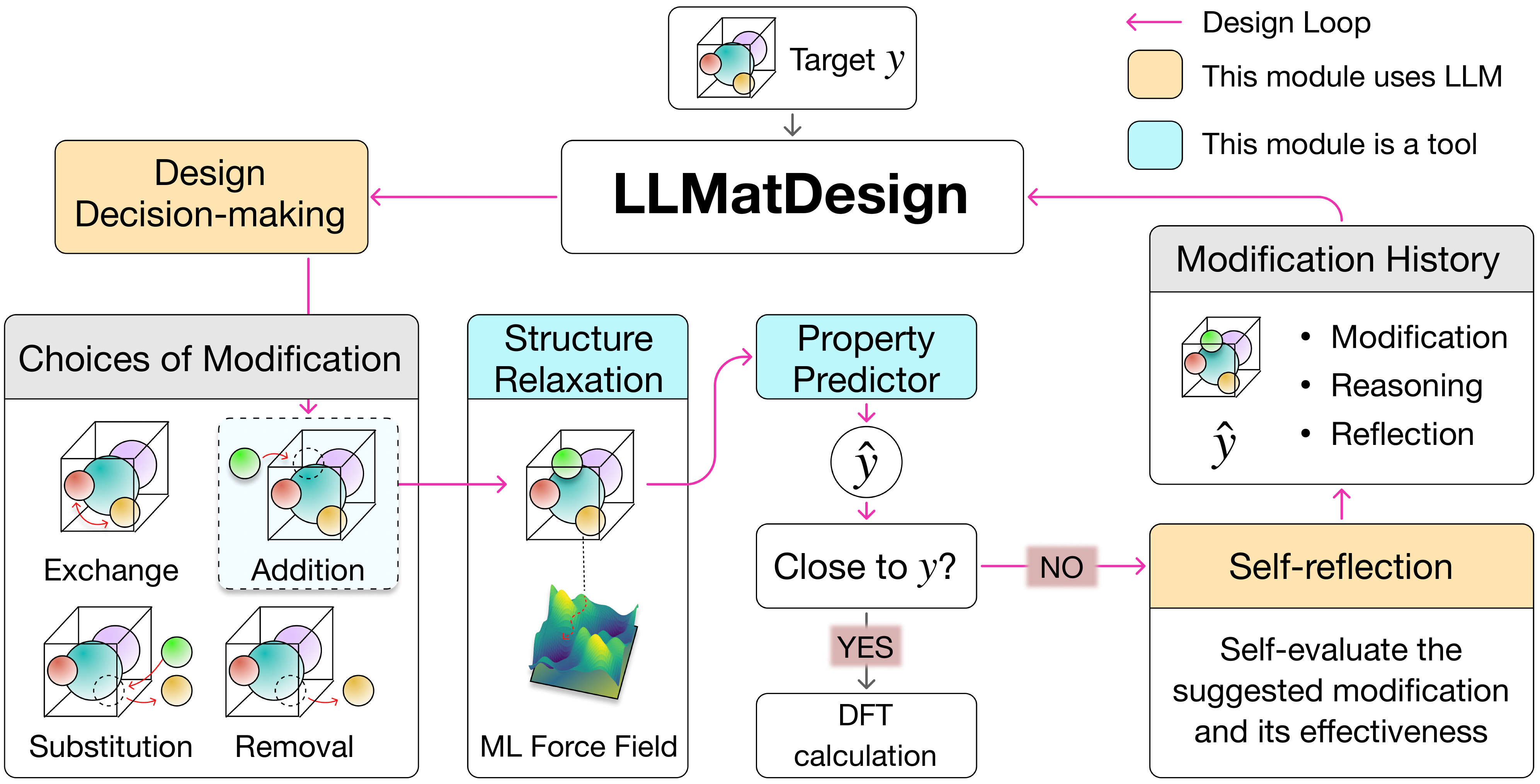}
    \caption{Overview of LLMatDesign. The discovery process with LLMatDesign begins with user-provided inputs of chemical composition and target property. It recommends modifications (addition, removal, substitution, or exchange), and uses machine learning tools for structure relaxation and property prediction. Driven by an LLM, this iterative process continues until the target property is achieved, with self-reflection on past modifications fed back into the decision-making process at each step.}
    \label{fig:overview}
\end{figure}

In this work, we present LLMatDesign (Fig. \ref{fig:overview}), a language-based framework for materials design powered by state-of-the-art LLMs. LLMatDesign is capable of interpreting human-provided instructions and design constraints, using computational tools for materials evaluation, and leveraging existing chemical knowledge and feedback to act as a highly effective autonomous materials design agent.
Unlike traditional methods that rely on explicit mathematical formulations and programmed solvers, LLMatDesign as an autonomous agent works with natural language directly, allowing it to quickly adapt to a diverse set of tasks, materials and target properties by simply modifying the prompt. In each step, LLMatDesign generates new designs of a material by choosing a modification of a starting material along with a corresponding hypothesis. It then applies the modification to the material and validates its property. Here, we use surrogate models as a stand-in for DFT to perform property validation, which can be readily replaced with any other computational or, potentially, experimental validation method. Following this, LLMatDesign reflects on the applied modification and its outcome. This reflection, along with the modified material and hypothesis, is then incorporated into the prompt in an iterative process. Moreover, LLMatDesign's flexibility allows incorporation of the entire modification history or user-defined requirements, offering even finer control over the discovery process. 


By utilizing state-of-the-art LLMs as chemical reasoning engines, LLMatDesign represents a novel framework for materials discovery which, unlike many current data-driven generative methods, eliminates the need for large training datasets derived from ab initio calculations. LLMatDesign's ability to interpret human instructions and incorporate design constraints enables rapid adaptation to new conditions, tasks, materials, and target properties via prompt modification---a flexibility that is often very difficult for current materials discovery methods such as those using generative models. More importantly, LLMatDesign's ability to generate hypothesis, evaluate outcomes, and self-reflect on past decisions in a closed-loop manner showcases the potential for a fully automated artificial intelligence (AI) agent for materials design in both a computational setting or towards robotic laboratories in the future.



\section{Results}\label{sec:background}

\subsection{LLMatDesign Framework}
LLMatDesign is a flexible framework powered by an LLM and empowered with the necessary tools to perform materials discovery. The discovery process with LLMatDesign begins by taking the chemical composition and property of a starting material, along with a target property value, as user-provided inputs. If a chemical composition is specified without an initial structure, LLMatDesign will automatically query the Materials Project \citep{jain2013commentary} database to retrieve the corresponding structure. If multiple candidates match the query, the structure with the lowest formation energy per atom is selected. LLMatDesign then intelligently recommends one of four possible modifications—addition, removal, substitution, or exchange—to the material's composition and structure to achieve the target value. 
Specifically, ``exchange" refers to swapping two elements within the material, while ``substitution" involves replacing one type of element with another. ``Removal" means eliminating a specific element from the material. In the case of ``addition," an atom of the suggested element is added to the unit cell of the material, with its position randomly determined. These four choices act as a proxy to physical processes in materials modification, such as doping or creating defects, and additional modification choices can also be readily added or removed as desired within the framework.

\noindent\begin{minipage}{\textwidth}
\begin{tcolorbox}[
    colback=white, colframe=black, coltitle=black, colbacktitle={rgb,255:red, 254; green, 229; blue, 178},
    title=LLMatDesign Prompt Template (GPT-4o), 
    fontupper=\footnotesize, 
    fontlower=\footnotesize
]
    I have a material and its {\color{red} \texttt{<property>}}. {\color{red}\texttt{<definition of property>}}.\newline
    
    ({\color{red}\texttt{<chemical composition>}}, {\color{red}\texttt{<property value>}})\newline
    
    Please propose a modification to the material that results in {\color{red}\texttt{<objective>}}. You can choose one of the four following modifications:

    \begin{enumerate}[noitemsep, leftmargin=*]
        \item exchange: exchange two elements in the material
        \item substitute: substitute one element in the material with another 
        \item remove: remove an element from the material 
        \item add: add an element to the material
    \end{enumerate}

    {\color{blue}\texttt{<additional constraints>}}\newline
    
    Your output should be a python dictionary of the following the format:\newline
    
    \{Hypothesis: \textdollar HYPOTHESIS, Modification: [\textdollar TYPE, \textdollar ELEMENT\_1, \textdollar ELEMENT\_2]\}. \newline
    
    Here are the requirements:
    \begin{enumerate}[noitemsep, leftmargin=*]
        \item \textdollar HYPOTHESIS should be your analysis and reason for choosing a modification 
        \item \textdollar TYPE should be the modification type; one of ``exchange", ``substitute", ``remove", ``add" 
        \item \textdollar ELEMENT should be the selected element type to be modified. For ``exchange" and ``substitute", two \textdollar ELEMENT placeholders are needed. For ``remove" and ``add", one \textdollar ELEMENT placeholder is needed. 
    \end{enumerate}

    {\color{blue}\texttt{<modification history>}}
\end{tcolorbox}
\captionof{figure}{Prompt template for LLMatDesign with GPT-4o. Text placeholders in red angular brackets are specific to the task given to LLMatDesign. Text placeholders in blue angular brackets are optional and can be omitted if not needed. For Gemini-1.0-pro's prompt template, see Appendix \ref{sec:appendix-prompts}.}\label{fig:basic_prompt}
\end{minipage}

\noindent\begin{minipage}{\textwidth}
        \begin{tcolorbox}[
            colback=white, colframe=black, coltitle=black, colbacktitle={rgb,255:red, 254; green, 229; blue, 178},
            title=Self-reflection Prompt Template, 
            fontupper=\footnotesize, 
            fontlower=\footnotesize
        ]
            After completing the following modification on {\color{red}\texttt{<previous composition>}}, we obtained {\color{red}\texttt{<current composition>}} and the {\color{red}\texttt{<property>}} changed from {\color{red}\texttt{<previous value>}} to {\color{red}\texttt{<current value>}}. Please write a brief post-action reflection on the modification, explaining how successful it was in achieving the {\color{red}\texttt{<objective>}} and the reasons for its success or failure: \newline
            
            {\color{red}\texttt{<hypothesis>, <modification>}}
        \end{tcolorbox}
        \captionof{figure}{Prompt template for self-reflection. Text placeholders in red angular brackets are specific to the task given to LLMatDesign.}
        \label{fig:self-reflection-prompt}
\end{minipage}

Alongside the proposed modification, LLMatDesign provides a hypothesis explaining why the suggested change could be beneficial. This hypothesis generated by the LLM provides a window into the reasoning behind its choices and provides a degree of interpretability which is not possible with traditional optimization algorithms. Next, LLMatDesign modifies the material based on the given suggestion, relaxes the structure using a machine learning force field (MLFF), and predicts its properties using a machine learning property predictor (MLPP). If the predicted property of the new material does not match the target value within a defined threshold, LLMatDesign then evaluates the effectiveness of the modification through a process called self-reflection where commentary is provided on the success of failure of the chosen modification.

After self-reflection, a modification history message is created. This message includes the modified chemical composition, the modification itself, the hypothesis behind the modification, and the self-reflection results. This history is then fed back into LLMatDesign, which enters the next design decision-making phase towards the goal of achieving the target property. The entire process repeats in a loop until termination conditions are met. Optionally, density functional theory (DFT) calculations can be performed on the final material. 

At the core of the entire workflow, LLMatDesign utilizes an LLM engine or agent which translates user-defined objectives into appropriate Materials Project API calls, drives the design decision-making process, and conducts self-reflection on previous decisions to enhance performance. In this work, we demonstrate the capabilities of LLMatDesign using two state-of-the-art LLMs: GPT-4o \citep{achiam2023gpt} and Gemini-1.0-pro \citep{team2023gemini}. However, the framework is model-agnostic and should function effectively with any capable LLMs. The overall architecture and algorithm of LLMatDesign is depicted in Fig. \ref{fig:overview} and Algo. \ref{alg:LLMatDesign} respectively. The modification and self-reflection prompt templates are shown in Fig. \ref{fig:basic_prompt} and \ref{fig:self-reflection-prompt} respectively.

\begin{algorithm}
\caption{LLMatDesign Algorithm}\label{alg:LLMatDesign}
\begin{algorithmic}
\State \textbf{Input:} $(x_0, y_0)$: chemical composition and property of the starting material. \par \hskip\algorithmicindent $y_{\text{target}}$: target property value to achieve. \par \hskip\algorithmicindent $\mathcal{M}:= \varnothing$: set of history messages, if any.
\State \textbf{Output:} $(x_i, y_i)$: chemical composition and property of the new material.
\For{$i = 1:N$} \Comment{$N$: maximum number of modifications}
    \State $s_i, h_i \gets \operatorname{LLM}(x_{i-1}, y_{i-1}, y_\text{target},\mathcal{M})$ \Comment{$s$: modification; $h$: hypothesis}
    \State $\tilde{x}_{i} \gets \texttt{perform\_modification}(x_{i-1}, s_i)$
    \State $x_{i}\gets \operatorname{MLFF}(\tilde{x}_{i})$ 
    \State $y_{i}\gets \operatorname{MLPP}(x_{i})$
    \If{$|y_{i} - y_{\text{target}}|/|y_{\text{target}}| \le \varepsilon$} \Comment{$\varepsilon$: error tolerance}
        \State \textbf{return} $(x_{i}, y_{i})$
    \EndIf
    \State $r_i \gets \operatorname{LLM}\left( x_{i-1}, x_{i}, y_{i-1}, y_{i}, s_i, h_i\right)$ \Comment{$r$: self-reflection}
    \State $m_i \gets \texttt{create\_history\_message}\left( s_i, h_i, r_i \right)$ \Comment{$m$: history message}
    \State $\mathcal{M}\gets \mathcal{M} \cup \{m_i\}$
\EndFor
\end{algorithmic}
\end{algorithm}





\subsection{Evaluation}
To evaluate the effectiveness of LLMatDesign, we performed a set of experiments with 10 starting materials randomly selected from the Materials Project \citep{chen2022universal}. Specifically, we focus on designing materials targeting two material properties and their corresponding objectives:
\begin{itemize}[leftmargin=*]
\item \textbf{Band gap} (eV): design a new material with a band gap of 1.4 eV.
\item \textbf{Formation energy per atom} (eV/atom): design a new material with the most negative formation energy possible.
\end{itemize}
The objective of achieving a band gap value of 1.4 eV is chosen as an example of designing an ideal photovoltaic material with a band gap within the range of 1–1.8 eV \citep{sutherland2020solar}, and the aim of obtaining the most negative formation energy requires LLMatDesign to suggest modifications that could result in more stable materials.

For the band gap experiments, we record the average number of modifications taken by LLMatDesign, with a maximum budget of up to 50 modifications. A 10\% tolerance of error to the target is used as the convergence criterion. For the formation energy experiments, a fixed budget of 50 modifications is used, and both the average and minimum formation energies are recorded. The experiment is then repeated 30 times for each starting material. We present results for two different LLM engines: Gemini-1.0-pro and GPT-4o. Within each LLM engine, two variants of experiments---\textit{history} and \textit{historyless}---are conducted to evaluate the impact of including the knowledge of prior modification history. All results are compared against a random baseline, where modifications to materials are randomly selected. The results for band gap and formation energy per atom are shown in Table \ref{tab:main-band-gaps} and Table \ref{tab:main-formation-energies} respectively. Note that self-reflection is included only for GPT-4o and not for Gemini-1.0-pro.

\begin{table}[ht]
\caption{LLMatDesign's performance in achieving a new material with a target band gap of 1.4 eV. Each experiment is repeated 30 times, and the average number of modifications taken to reach the target value is recorded.}
\label{tab:main-band-gaps}
\footnotesize
\begin{center}
\renewcommand\cellalign{c}
\setcellgapes{3pt}\makegapedcells
\begin{adjustbox}{max width=\textwidth}
\begin{tabular}{|c|ccccc|ccccc|}
\hline
\multirow{3}{*}{Starting Material} & \multicolumn{5}{c}{Average \# of Modifications}                                                         & \multicolumn{5}{c|}{Average Final Band Gap (eV)}                         \\ \cline{2-11} 
                                   & \multicolumn{2}{c}{\underline{Gemini-1.0-pro}} & \multicolumn{2}{c}{\underline{GPT-4o}}      & {\underline{Random}} & \multicolumn{2}{c}{\underline{Gemini-1.0-pro}} & \multicolumn{2}{c}{\underline{GPT-4o}}      &{\underline{Random}} \\
                                   & History        & Historyless       & History        & Historyless    &                         & History           & Historyless    & History        & Historyless    &                         \\ [2pt]\hline
$\text{BaV}_2\text{Ni}_2\text{O}_8$                          & 17.7              & {14.4}           & 17.7 & 30.4         & \multicolumn{1}{c|}{22.4}   & 1.23           & 1.42              & {1.39}       & 1.89          & 1.12   \\
$\text{CdCu}_2\text{GeS}_4$                          & 11.1              & 13.4           & {3.3}  & 9.5          & \multicolumn{1}{c|}{28.7}   & {1.41}  & {1.39}     & 1.44       & 1.38          & 1.01   \\
$\text{CeAlO}_3$                             & 14.3              & 15.1           & {7.4}  & 16.9         & \multicolumn{1}{c|}{26.7}   & 1.42           & {1.39}     & {1.41 }  & 1.68          & 1.21   \\
$\text{Co}_2\text{TiO}_4$                            & 8.8               & 13.1           & 5.5           & {1.6} & \multicolumn{1}{c|}{29.7}   & {1.40}  & 1.30              & 1.36       & 1.42          & 1.02   \\
$\text{ErNi}_2\text{Ge}_2$                           & 26.8              & 24.8           & {19.3} & 47.6         & \multicolumn{1}{c|}{31.8}   & 1.18           & 1.26              & {1.36}       & 0.43          & 0.90   \\
$\text{Ga}_2\text{O}_3$                              & {10.3}     & 12.3           & 12.7          & 37.7         & \multicolumn{1}{c|}{32.8}   & 1.34           & {1.38}              & 1.36       & 1.76          & 0.87   \\
$\text{Li}_2\text{CaSiO}_4$                          & 15.7              & 20.5           & {14.3} & 29.3         & \multicolumn{1}{c|}{27.4}   & 1.36           & 1.37              & {1.41}       & 1.81          & 1.09   \\
LiSiNO                             & 12.4              & 10.4           & 4.1           & {2.8} & \multicolumn{1}{c|}{27.4}   & 1.38           & {1.39}              & {1.39}       & 1.50          & 1.09   \\
$\text{Na}_2\text{ZnGeO}_4$                          & 13.0              & 15.0           & {11.5} & 49.4         & \multicolumn{1}{c|}{22.9}   & {1.40}           & 1.39              & 1.39       & 2.35          & 1.15   \\
$\text{SrTiO}_3$                             & {7.2}      & 8.8            & 12.0          & 40.6         & \multicolumn{1}{c|}{24.3}   & 1.42           & {1.41}              & 1.45       & 1.64          & 1.11   \\ [2pt] \hline
Avg.                               & 13.7              & 14.8           & \textbf{10.8} & 26.6         & \multicolumn{1}{c|}{27.4}   & 1.35           & 1.37              & \textbf{1.39}       & 1.59          & 1.06  \\ [2pt] \hline
\end{tabular}
\end{adjustbox}
\end{center}
\end{table}

\begin{table}[ht]
\caption{LLMatDesign's performance in achieving a new material with a as low as possible formation energy per atom. Each experiment consists of 50 modifications, and is repeated 30 times.}
\label{tab:main-formation-energies}
\footnotesize
\begin{center}
\renewcommand\cellalign{c}
\setcellgapes{3pt}\makegapedcells
\begin{adjustbox}{max width=\textwidth}
\begin{tabular}{|c|ccccc|ccccc|}
\hline
\multirow{3}{*}{Starting Material} & \multicolumn{5}{c}{Average Formation Energy (eV/atom)}                                                                    & \multicolumn{5}{c|}{Minimum Formation Energy (eV/atom)}                                                                   \\ \cline{2-11} 
                                   & \multicolumn{2}{c}{\underline{Gemini-1.0-pro}} & \multicolumn{2}{c}{\underline{GPT-4o}}      & {\underline{Random}} & \multicolumn{2}{c}{\underline{Gemini-1.0-pro}} & \multicolumn{2}{c}{\underline{GPT-4o}}      &{\underline{Random}} \\
                                   & History        & Historyless       & History        & Historyless    &                         & History           & Historyless    & History        & Historyless    &                         \\ [2pt]\hline
$\text{BaV}_2\text{Ni}_2\text{O}_8$                          & -0.80          & -0.20             & -2.45          & {-2.50}  & -0.12                   & -2.69             & -2.30          & {-2.91} & -2.74          & -1.99                   \\
$\text{CdCu}_2\text{GeS}_4$                          & -0.19          & 0.11              & {-1.05} & -0.61          & 0.29                    & -1.31             & -1.59          & {-1.61} & -0.72          & -1.37                   \\
$\text{CeAlO}_3$                             & -0.77          & -0.28             & {-2.79}  & -2.24          & -0.04                   & -3.44             & -3.22          & {-3.73} & {-3.73}          & -2.50                   \\
$\text{}\text{TiO}_4$                            & -0.39          & 0.03              & {-1.57} & -1.49          & 0.0                     & {-2.64}    & -2.08          & -2.48          & -2.10          & -1.80                   \\
$\text{ErNi}_2\text{Ge}_2$                           & -0.02          & -0.19             & -0.54         & {-0.74} & -0.11                   & -0.96             & -1.71          & -0.94          & {-1.57} & -1.40                   \\
$\text{Ga}_2\text{O}_3$                              & -0.19          & -0.12             & -1.61          & {-2.07} & -0.16                   & -2.05             & -1.82          & {-3.31} & -3.29          & -1.67                   \\
$\text{Li}_2\text{CaSiO}_4$                          & -0.77          & -0.41             & -2.30          & {-2.69} & -0.20                   & -2.94             & -2.60          & {-3.13} & -2.98          & -2.27                   \\
LiSiNO                             & -0.38          & -0.19             & {-1.75} & -1.54          & -0.15                   & -2.01             & -2.01          & {-2.60} & -1.72          & -1.75                   \\
$\text{Na}_2\text{ZnGeO}_4$                          & -0.79          & -0.25             & {-2.62} & -2.52          & 0.05                    & -2.48             & -2.34          & {-2.87} & -2.55          & -1.85                   \\ 
$\text{SrTiO}_3$                             & -1.26          & -0.23             & -3.01          & {-3.54} & -0.02                   & -3.40             & -3.09          & {-3.65} & -3.57          & -2.38                   \\ [2pt] \hline 
Avg.                               & -0.56          & -0.17             & -1.97          & \textbf{-1.99} & -0.05                   & -2.39             & -2.28          & \textbf{-2.72} & -2.50          & -1.90 \\ [2pt] \hline
\end{tabular}
\end{adjustbox}
\end{center}
\end{table}

\begin{figure}[h]
    \centering
    \includegraphics[width=0.95\textwidth]{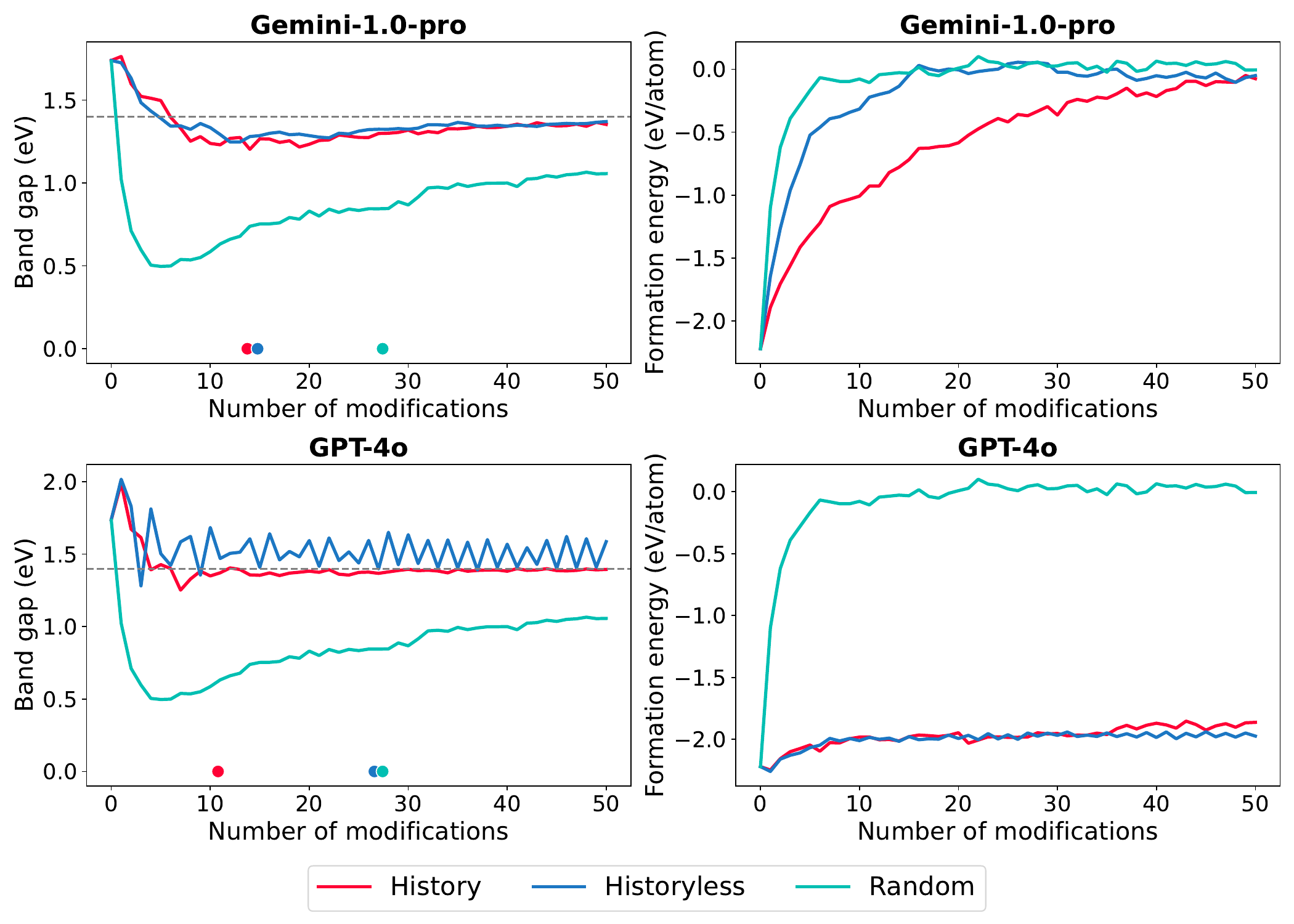}
    \caption{Average band gaps and formation energies over 50 modifications. The grey horizontal line indicates the target band gap of 1.4 eV. The colored dots on the x-axis indicate the average number of modifications taken for each method to reach the target. For formation energy, the goal is to achieve the lowest possible value.}
    \label{fig:avg_convergence}
\end{figure}

We observe that GPT-4o with past modification history performs the best in achieving the target band gap value of 1.4 eV, requiring an average of 10.8 modifications (Table \ref{tab:main-band-gaps}). In comparison, Gemini-1.0-pro with history takes an average of 13.7 modifications. Both methods signifcantly outperform the baseline, whic requires 27.4 modifications. Adding modification history to subsequent prompts allows the LLMs to converge to the target more quickly, as both Gemini-1.0-pro and GPT-4o with modification history outperform their historyless counterparts. Notably, the performance gap between the history and historyless variants is smaller for Gemini-1.0-pro than for GPT-4o. From a closer inspection of the modification paths of GPT-4o without history, we find that GPT-4o often alternates between a few of the same modifications until reaching the maximum number of allowed iterations (see Fig. \ref{fig:avg_convergence}). For the two starting materials where GPT-4o without history performs the best ($\text{Co}_2\text{TiO}_4$ and $\text{SrTiO}_3$), the final materials frequently converge to identical composition by following the same modification sequence. This indicates a lack of diversity in the newly generated materials when no history is included in LLMatDesign's iterative loop. In addition, GPT-4o with history achieves the best final band gap value, averaging 1.39 eV, followed by Gemini-1.0-pro at 1.35 eV, and random at 1.06 eV. 

LLMatDesign's superior performance is also apparent when finding new materials with the lowest formation energy per atom (Table \ref{tab:main-formation-energies}), consistently outperforming the random baseline. Specifically, both the history and historyless variants of GPT-4o achieve the lowest average formation energies, with $-1.97$ eV/atom and $-1.99$ eV/atom, respectively. GPT-4o with history also achieves the lowest minimum formation energy per atom at $-2.72$ eV/atom. Interestingly, while the minimum formation energy per atom values achieved by Gemini-1.0-pro are close to that of GPT-4o, its average formation energy per atom values are significantly higher, indicating that it struggles to consistently suggest chemically stable modifications for the materials. Nonetheless, Gemini-1.0-pro still noticeably outperforms the baseline. In Fig. \ref{fig:band_vis} and \ref{fig:formation_vis}, we visualize 20 materials discovered by LLMatDesign for the band gap and formation energy tasks, respectively. These materials are obtained from the first run of all 10 starting materials. For the band gap task, the final materials are selected. For the formation energy task, the materials with the lowest formation energy per atom are chosen. 

In Fig. \ref{fig:avg_convergence}, we plot the band gaps and formation energies per atom over 50 modifications, averaged across 10 starting materials. The target band gap of 1.4 eV is indicated by the grey horizontal line. Both history and historyless variants of Gemini-1.0-pro and GPT-4o demonstrate quick convergence to the target band gap. However, the GPT-4o historyless variant exhibits zig-zag oscillations in band gap values as modifications increase. This occurs because, without historical information, GPT-4o tends to oscillate between a few of the same moves, causing the band gap to fluctuate without improving. In contrast, the random baseline fails to converge to 1.4 eV within the maximum allowed 50 modifications. For formation energy, our findings indicate that GPT-4o is consistently able to suggest modifications which keep formation energy low on average around $-2$ eV/atom, though Gemini-1.0-pro struggles to do so despite being able to obtain a low minimum formation energy. Notably, neither GPT-4o nor Gemini-1.0-pro are able to beat the formation energy of the starting materials, likely due to the the fact that these materials are already at or near the lowest energy states.


\begin{figure}[h]
    \centering
    \begin{minipage}[b]{0.32\textwidth}
        \centering
        \includegraphics[width=\textwidth]{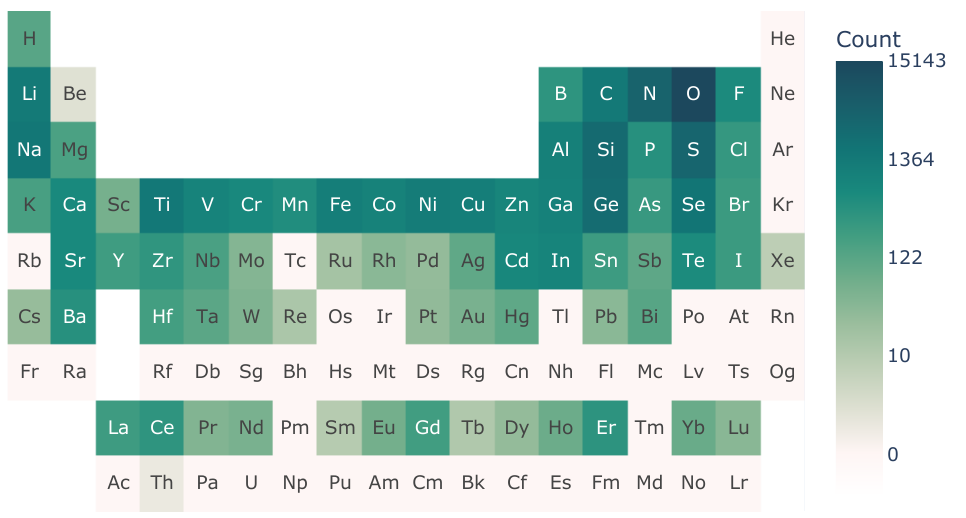}
        \captionsetup{font=small}
        \caption*{BG: Gemini-1.0-pro with history}
        \label{fig:periodic-BG-history-gemini-1.0-pro}
    \end{minipage}
    \hfill
    \begin{minipage}[b]{0.32\textwidth}
        \centering
        \includegraphics[width=\textwidth]{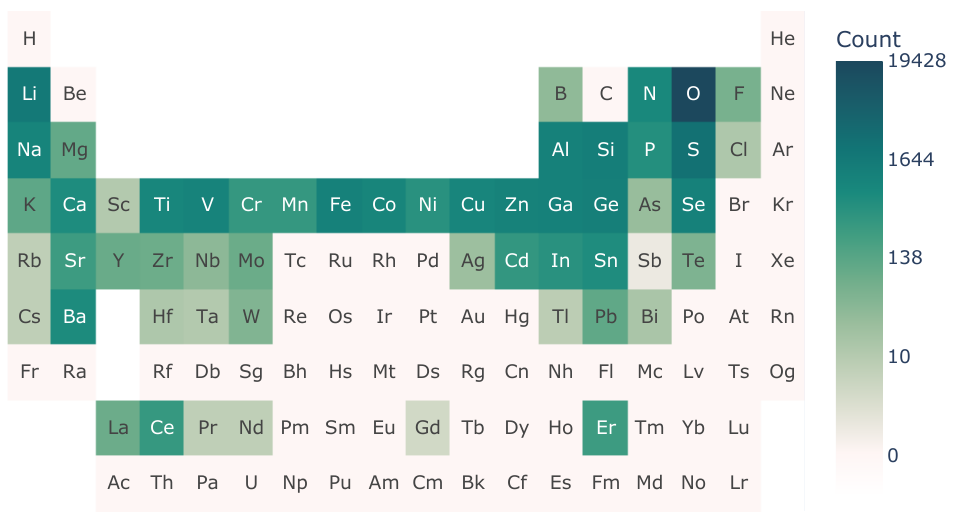}
        \captionsetup{font=small}
        \caption*{BG: GPT-4o with history}
        \label{fig:periodic-BG-history-GPT-4o}
    \end{minipage}
    \hfill
    \begin{minipage}[b]{0.32\textwidth}
        \centering
        \includegraphics[width=\textwidth]{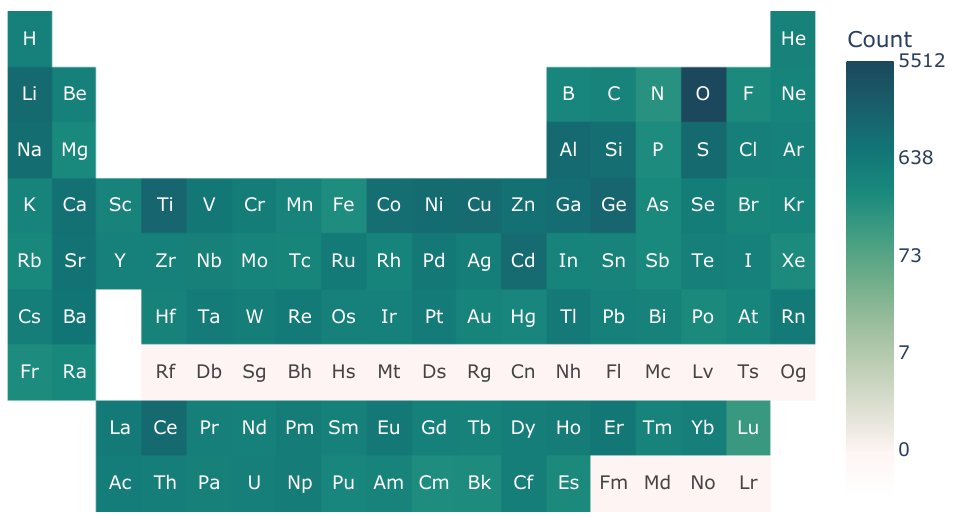}
        \captionsetup{font=small}
        \caption*{BG: Random}
        \label{fig:periodic-BG-random}
    \end{minipage}
    \begin{minipage}[b]{0.32\textwidth}
        \centering
        \includegraphics[width=\textwidth]{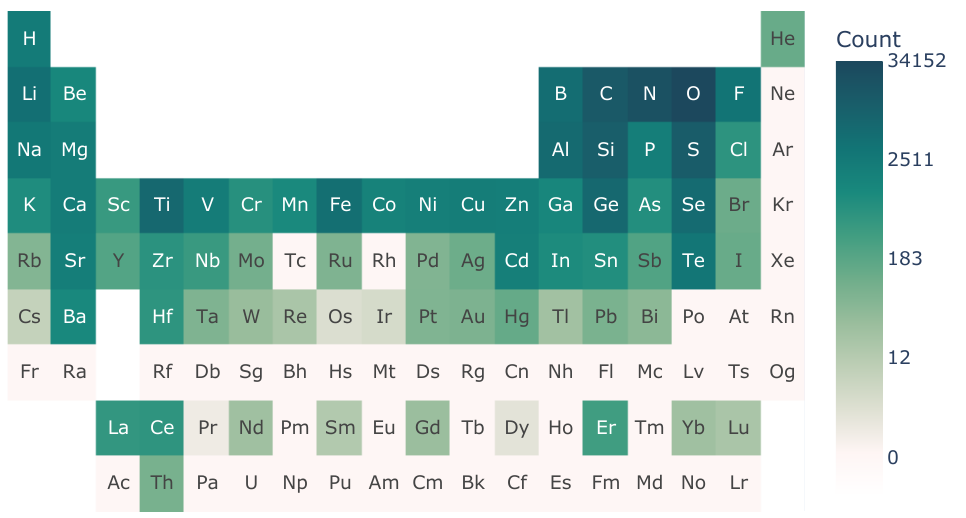}
        \captionsetup{font=small}
        \caption*{FE: Gemini-1.0-pro with history}
        \label{fig:periodic-FE-history-gemini-1.0-pro}
    \end{minipage}
    \hfill
    \begin{minipage}[b]{0.32\textwidth}
        \centering
        \includegraphics[width=\textwidth]{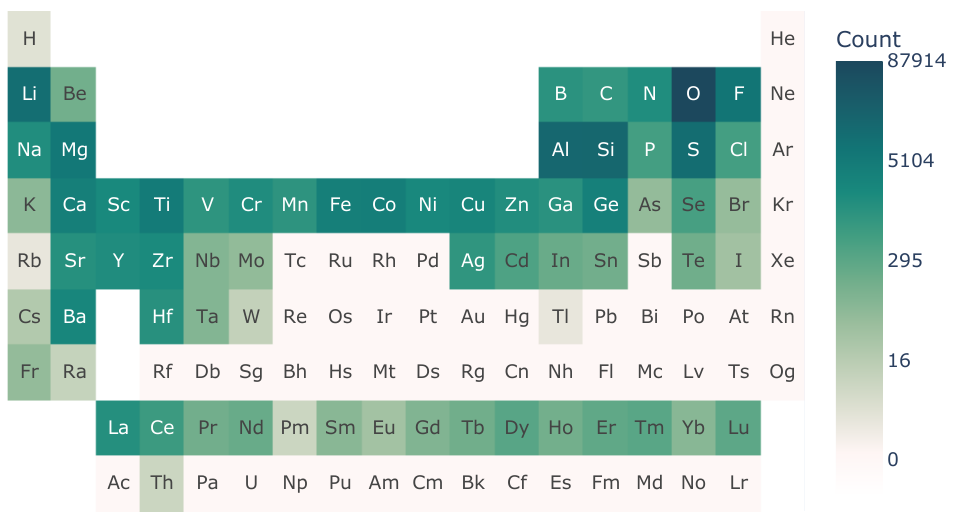}
        \captionsetup{font=small}
        \caption*{FE: GPT-4o with history}
        \label{fig:periodic-FE-history-GPT-4o}
    \end{minipage}
    \hfill
    \begin{minipage}[b]{0.32\textwidth}
        \centering
        \includegraphics[width=\textwidth]{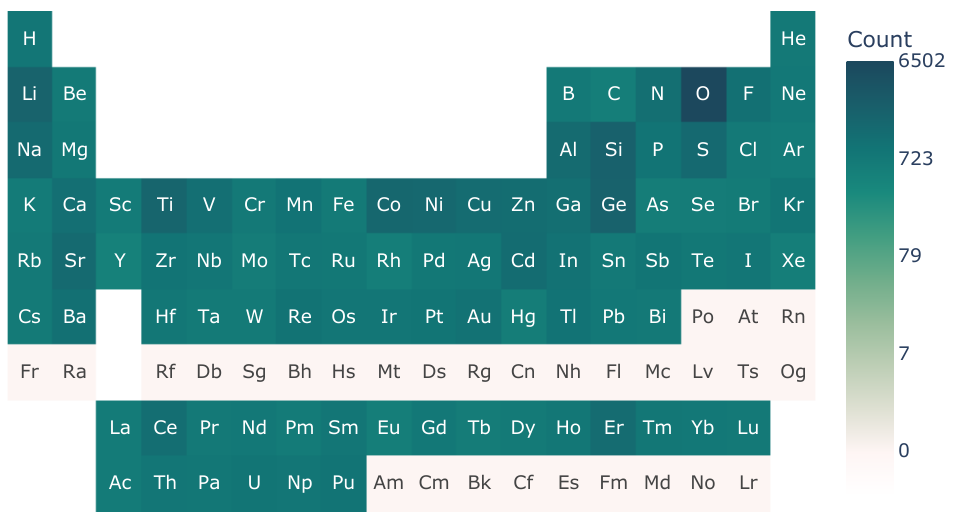}
        \captionsetup{font=small}
        \caption*{FE: Random}
        \label{fig:periodic-FE-random}
    \end{minipage}
    
    \caption{Heatmaps of element frequencies in band gap (BG) and formation energy (FE) tasks. The periodic table is color-coded to indicate the frequency of each element's occurrence in all modified materials (both intermediate and final) across all runs and starting materials. Darker colors represent higher frequencies, while lighter colors denote lower frequencies or absence. The visualization employs log-scaling to effectively highlight the distribution and prevalence of elements.}
    \label{fig:periodic_tables_heatmaps}
\end{figure}

Fig. \ref{fig:periodic_tables_heatmaps} presents heatmaps over the periodic table displaying the element occurrences in the modifications for both the band gap and formation energy tasks, which reveal additional insights into the reason for the good performance for LLM-driven design. The number of occurrences of each element is collected across all runs and starting materials. In the heatmaps for the random baseline, all elements are chosen at nearly uniform frequencies. This result is to be expected, as the random algorithm samples elements with atomic numbers up to 99 uniformly. Meanwhile, in the heatmaps for the LLM cases, there is a clear distribution towards certain elements, mostly focusing on elements within the first four rows of the periodic table and avoiding noble metals and Actinides. Both LLM models share similar distributions, such as a preference for elements like oxygen, however Gemini-1.0-pro's suggestions appear to exhibit a greater element diversity compared to GPT-4o, including some of the transition metals. With Gemini-1.0-pro, we also occasionally observe modifications suggested by the LLM that include noble gases, which is not chemically feasible due to their inert nature. With GPT-4o, this does not occur (see Fig. \ref{fig:periodic_tables_heatmaps_historyless}). Regardless, both LLM models are able to consistently suggest chemically viable elements for modification, which is akin to how a human expert would make similar choices based on chemical intuition or from past examples in the literature.

\begin{figure}[H]
    \centering
    \includegraphics[width=0.65\textwidth]{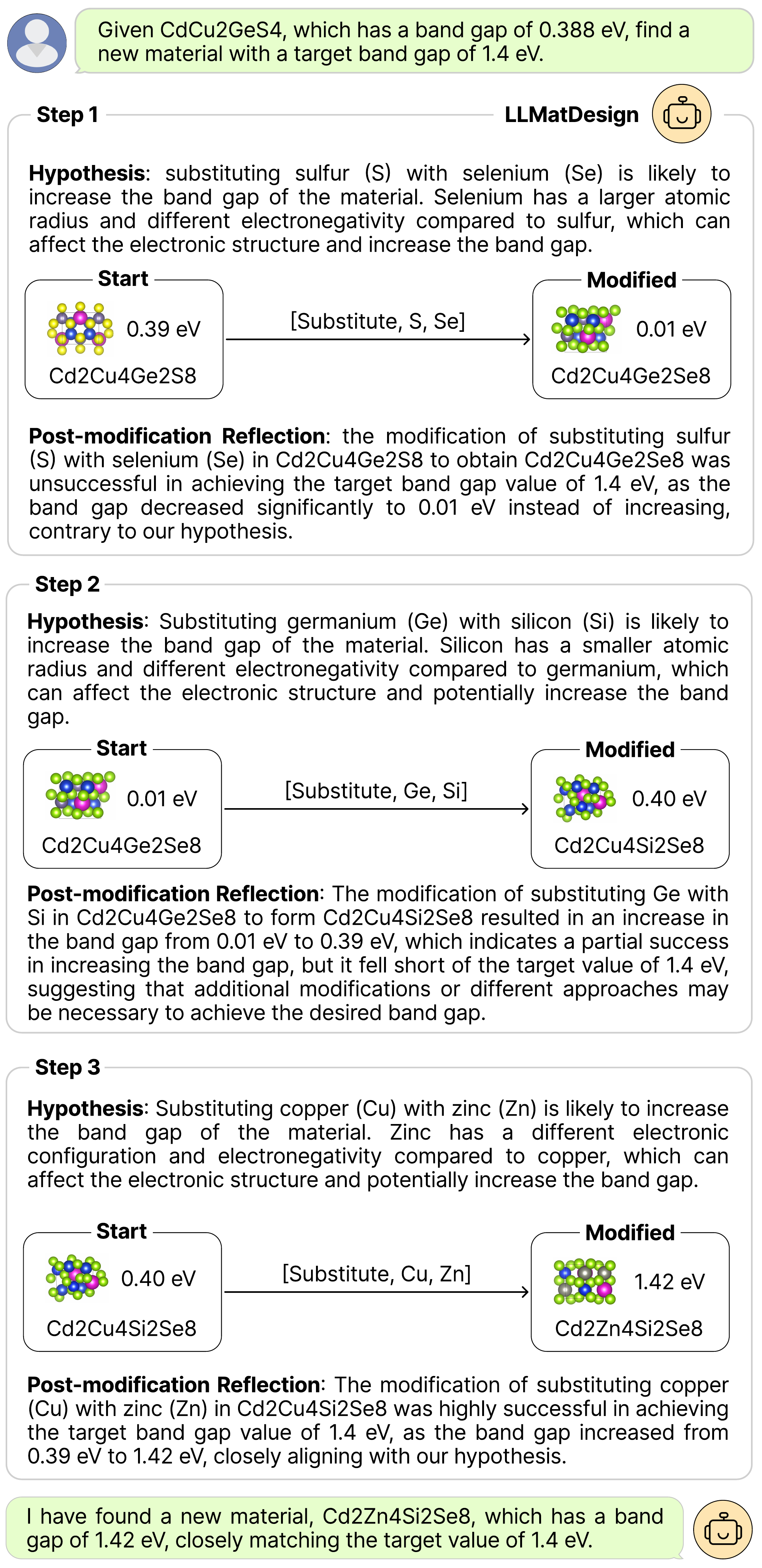}
    \caption{Example of LLMatDesign with GPT-4o on the task of modifying the starting material $\text{CdCu}_2\text{GeS}_4$ to achieve a band gap of 1.40 eV. The starting material is retrieved from the Materials Project with chemical formula $\text{Cd}_2\text{Cu}_4\text{Ge}_2\text{S}_8$.}
    \label{fig:example}
\end{figure}

In Fig. \ref{fig:example}, we present an example of the full process whereby LLMatDesign successfully completes a design task to achieve a band gap of 1.40 eV. In the first step, LLMatDesign suggests modifying the starting material $\text{CdCu}_2\text{GeS}_4$ by substituting S with Se, given the hypothesis that increasing atomic radius and changing the electronegativity can alter the band gap. Upon modification, the new material $\text{CdCu}_2\text{GeSe}_4$ was found to have an even smaller band gap, which is contrary to the desired effect as noted by the reflection. This history is included in the second step of modification, whereby LLMatDesign suggests a subsequent modification of Ge with Si, which increases the gap. The reflection notes a partial success is achieved, but is still not enough to reach the target, whereupon a third step is taken. In the third step, Cu is substituted with Zn, which finally achieves the desired band gap within an acceptable threshold, ending the process. From this example, we can observe the LLM is successful at 1) recognizing differences in element properties (i.e. Se having a larger atomic radius than S), 2) highlighting these properties as being relevant to the design task (i.e. atomic radius, electronegativity, and electronic configuration affecting the band hap), 3) and recognizing whether a modification is successful and the degree of success in the reflection. We will show in the subsequent section that it is this reasoning and reflection process which has a significant impact on its success.

In the final step of the design process, a DFT calculation is performed to validate the material's properties which were obtained from a ML surrogate model. Here, we use DFT to compute the formation energy of the minimum energy structures in all 30 runs for each 10 starting materials obtained with GPT-4o and random sampling. On average, structures generated by LLMatDesign using GPT-4o with history achieved a formation energy of -2.32 eV/atom with a job completion rate of 73.3\%. In comparison, the random baseline obtained an average formation energy of -1.51 eV/atom, with a significantly lower job completion rate of 40.0\% (see Fig. \ref{tab:dft-results-formation-energies}). The lower completion rate is likely due to a much higher likelihood of unreasonable compositions and configurations obtained from random sampling which cause the DFT calculations to fail. These results also follow the same relative trends in Table \ref{tab:main-formation-energies}. In a situation where DFT is directly used in each iteration rather than a surrogate model, this step can be omitted, or potentially the materials can be validated further with experiments. 

\subsection{Self-reflection}
To quantify the effect of self-reflection on the performance of LLMatDesign, we conduct band gap experiments using GPT-4o and the same set of 10 starting materials, where we aim to find a new material with a target band gap of 1.4 eV. Like with the history variant, past modifications are incorporated into the prompting loop. However, in this case, self-reflection is omitted completely. In other words, the history message only includes the modification and hypothesis pairs (see Algo. \ref{alg:LLMatDesign}). The results from these experiments are shown in Table \ref{tab:self-reflection}. As previously discussed, GPT-4o with history achieves an average of 10.8 modifications, while GPT-4o without history requires 26.6 modifications. In comparison, GPT-4o with history but without self-reflection now needs an average of 23.4 modifications, which is over twice as many compared to including self-reflection. These results suggest that self-reflection, which involves the LLM evaluating and reasoning through its previous design choices, plays a crucial role in enhancing the efficiency of LLMatDesign in achieving the given objective. 

\begin{table}[ht]
\caption{LLMatDesign with and without self-reflection. GPT-4o is used as the LLM engine.}
\label{tab:self-reflection}
\footnotesize
\begin{center}
\renewcommand\cellalign{c}
\setcellgapes{3pt}\makegapedcells
\begin{adjustbox}{max width=\textwidth}
\begin{tabular}{|c|ccc|}
        \hline
        \multirow{3}{*}{Starting Material} 
        & \multicolumn{3}{c|}{Average \# of Modifications} \\ \cline{2-4} 
                                           & History & Historyless & History without reflection \\ \hline
        $\text{BaV}_2\text{Ni}_2\text{O}_8$                          & 17.7    & 30.4        & 45.1                       \\
        $\text{CdCu}_2\text{GeS}_4$                          & 3.3     & 9.5         & 5.0                        \\
        $\text{CeAlO}_3$                             & 7.4     & 16.9        & 27.6                       \\
        $\text{Co}_2\text{TiO}_4$                            & 5.5     & 1.6         & 7.9                        \\
        $\text{ErNi}_2\text{Ge}_2$                           & 19.3    & 47.6        & 31.0                       \\
        $\text{Ga}_2\text{O}_3$                              & 12.7    & 37.7        & 13.1                       \\
        $\text{Li}_2\text{CaSiO}_4$                          & 14.3    & 29.3        & 31.4                       \\
        LiSiNO                             & 4.1     & 2.8         & 5.1                        \\
        $\text{Na}_2\text{ZnGeO}_4$                          & 11.5    & 49.4        & 31.5                       \\
        $\text{SrTiO}_3$                             & 12.0    & 40.6        & 36.7                       \\[2pt] \hline
        Avg.                               & \textbf{10.8}    & 26.6        & 23.4                       \\ [2pt] \hline
        \end{tabular}
\end{adjustbox}
\end{center}
\end{table}

\subsection{Prompting}
Well-crafted prompts are essential for eliciting accurate and useful responses from LLMs. While the base prompt template, shown in Fig. \ref{fig:basic_prompt}, works as intended, we subsequently show that optimizing this prompt can improve the performance of LLMatDesign even further. To this end, we develop two additional prompt templates in a non-exhaustive demonstration. The first template, termed \textit{GPT-4o Refined}, is an enhancement of the original prompt (Fig. \ref{fig:basic_prompt}) created by GPT-4o itself. This refinement includes rephrasing and reformatting parts of the original prompt and appending the following sentence: ``Take a deep breath and work on this problem step-by-step. Your thoughtful and detailed analysis is highly appreciated." The second template, named \textit{Persona}, mirrors the original prompt but incorporates the persona of a materials specialist. Specifically, it begins with a declaration that the LLM is a materials design expert working on developing new materials with specific properties. Detailed descriptions of these prompt templates are provided in Appendix \ref{sec:appendix-prompts}. 


We conduct the same experiments on the band gap task using GPT-4o as the LLM engine for LLMatDesign across all 10 starting materials. The results, shown in Table \ref{tab:prompt-design}, indicate that both the GPT-4o Refined and Persona prompt templates outperform the GPT-4o with history, with the GPT-4o Refined template achieving the best performance, requiring an average of only 8.69 modifications to complete the task. The improvement over the original prompt template indicates that careful prompt optimization can positively enhance the efficiency and accuracy of LLM-directed materials discovery frameworks, and that this process can even be performed by the LLM itself. This is a particularly intriguing discovery as it hints towards an unprecedented level of autonomy which can be enabled by LLMs, whereby the prompts and instructions in the framework can be continuously tuned in an automated manner with minimal human intervention. 

\begin{table}[ht]
\caption{LLMatDesign with different prompts. GPT-4o is used as the LLM engine.}
\label{tab:prompt-design}
\footnotesize
\begin{center}
\renewcommand\cellalign{c}
\setcellgapes{3pt}\makegapedcells
\begin{adjustbox}{max width=\textwidth}
\begin{tabular}{|c|ccc|}
\hline
\multirow{3}{*}{Starting Material} & \multicolumn{3}{c|}{Average \# of Modifications} \\ \cline{2-4} 
                                   & History       & GPT-4o Refined       & Persona      \\ \hline
$\text{BaV}_2\text{Ni}_2\text{O}_8$                          & 17.7         & 13.4                  & 9.6          \\
$\text{CdCu}_2\text{GeS}_4$                          & 3.3          & 3.1                   & 5.3          \\
$\text{CeAlO}_3$                             & 7.4           & 7.2                   & 8.7          \\
$\text{Co}_2\text{TiO}_4$                            & 5.5          & 8.9                   & 11.9         \\
$\text{ErNi}_2\text{Ge}_2$                           & 19.3          & 11.9                  & 11.7         \\
$\text{Ga}_2\text{O}_3$                              & 12.7         & 8.4                   & 8.3          \\
$\text{Li}_2\text{CaSiO}_4$                          & 14.3         & 11.6                  & 11.9         \\
LiSiNO                             & 4.1           & 5.6                   & 1.0            \\
$\text{Na}_2\text{ZnGeO}_4$                          & 11.5         & 6.9                   & 8.8          \\
$\text{SrTiO}_3$                             & 12.0         & 9.9                   & 13.9         \\ [2pt]\hline
Avg.                               & 10.8         & \textbf{8.69}                  & 9.11         \\ [2pt]\hline
\end{tabular}
\end{adjustbox}
\end{center}
\end{table}

\subsection{Constrained Materials Design}
Materials discovery with constraints ensures scientific, economic, and political viability. For instance, avoiding the use of rare earth metals can reduce dependency on limited and expensive resources, mitigate supply chain risks, and align with environmental and ethical standards. To this end, we evaluate LLMatDesign under three constraints limiting its action space. Experiments are conducted on the band gap task using the starting material $\text{SrTiO}_3$ with GPT-4o to test whether these constraints are obeyed. Like before, each experiment is repeated 30 times, and the percentage of modifications adhering strictly to the constraints is calculated across all runs. As shown in Table \ref{tab:constraints}, LLMatDesign perfectly adheres to the constraints of ``do not use Ba or Ca" and ``do not modify Sr," achieving 100\% compliant modifications. For the constraint ``do not have more than 4 distinct elements," only 4 out of 509 modifications by LLMatDesign include 5 distinct elements, resulting in a high compliance rate of 99.02\%. These results demonstrate LLMatDesign's robust capability in adhering to predefined constraints as described by natural language, an advantage unique to LLM-driven design. 

\begin{table}[ht]
\caption{LLMatDesign with different constraints on $\text{SrTiO}_3$. }
\label{tab:constraints}
\footnotesize
\begin{center}
\renewcommand\cellalign{c}
\setcellgapes{3pt}\makegapedcells
\begin{adjustbox}{max width=\textwidth}
\begin{tabular}{lc}
\hline
Constraint                                & \% compliant modifications \\ \hline
Do not use Ba or Ca                       & 100                             \\
Do not modify Sr                          & 100                             \\
Do not have more than 4 distinct elements & 99.02                           \\ \hline
\end{tabular}
\end{adjustbox}
\end{center}
\end{table}

\subsection{Further Discussion}

Through extensive experiments, we find LLMatDesign consistently outperforms baselines by a significant margin, demonstrating the viability of using LLM-based autonomous agents for materials discovery tasks under a limited budget. While the random baseline uniformly samples from a set of elements for modification (see Fig. \ref{fig:periodic_tables_heatmaps}), LLMatDesign, whether utilizing GPT-4o or Gemini-1.0-pro, exhibits inherent chemical knowledge, enabling it to provide chemically meaningful suggestions. Furthermore, GPT-4o accurately recognizes periodic trends such as atomic radius and electronegativity in its hypotheses and self-reflections in guiding its decisions. In contrast, Gemini-1.0-pro is more prone to errors in this regard, likely due to it being a less robust LLM. Further experiments also show the critical role of self-reflection in the performance of the LLM. This indicates that by reviewing and learning from its previous decisions, LLMatDesign can refine its future suggestions more effectively. This iterative learning process helps the model understand the implications of its modifications better, leading to quicker convergence. In general, it is evident that there are more complex underpinnings behind the remarkable effectiveness of LLM-driven design than simply predicting most likely outcomes. 

This work also demonstrates the \textit{lower-bound} capabilities of LLM-based design, which is performed without further fine-tuning in a zero-shot manner. A natural extension of this approach would be to further train LLMs on chemical and materials knowledge, such as those obtained from literature articles. In the future, it would be highly desirable for a chemically fine-tuned to provide more insightful hypotheses and explanations, and even refer to specific references of prior published experiments to support them. These capabilities can potentially be within reach given the growing prevalence of powerful open-source LLMs and parameter-efficient fine-tuning.

In the current examples, LLMatDesign comes up with new materials designs from a limited set of modifications on the composition of a material. Nonetheless, this framework is general and can include more complex modifications which act not only on the composition space but also the structure space. Future work in this direction will focus on incorporating structural information when describing the material being modified, and also suggest modifications which directly act on the positions and lattice of the crystal structure. To this end, recent advances in multimodal LLMs can be applied here, where the atomic structure is considered to be an additional modality to be encoded in addition to the text modality.

\section{Conclusion}
In this work, we present LLMatDesign, a novel materials design framework powered by state-of-the-art LLMs that works directly with user-defined design requirements and constraints in natural language. It integrates computational tools for structure relaxation and property evaluation, incorporates internal chemical knowledge, and learns from previous iterations to function as an automated material design framework with high efficiency.
Additionally, LLMatDesign quickly adapt to different tasks, target properties and design constraints by simply modifying the prompt. In our experiments, LLMatDesign consistently outperforms the baseline, demonstrating the effectiveness of the framework in developing new materials. Our work highlights the potential for fully automated AI-driven materials discovery that can be seamlessly integrated into autonomous laboratories in the future.

\section{Methods}
\subsection{Large Language Models}
Large language models (LLMs) are a class of machine learning models built on the transformer architecture \citep{vaswani2017attention}. By training on vast amounts of text data, these models can understand and generate text in a human-like manner. In this work, GPT-4o \citep{achiam2023gpt} refers to OpenAI's \texttt{gpt-4o} model, which has a context length of 128K and a knowledge cutoff date of October 2023. Gemini-1.0-pro \citep{team2023gemini} refers to Google's LLM with the same name, featuring a context length of 32K.

\subsection{Machine Learning Force Field}
Machine learning force fields (MLFFs) represent a significant advancement in computational chemistry and materials science. By utilizing state-of-the-art machine learning models and training on extensive datasets of atomic structures with energies, forces, and stresses, MLFFs can achieve high accuracy in predicting these properties, often rivaling ab initio methods such as density functional theory (DFT) \citep{ko2023recent}. More importantly, MLFFs provide these high-accuracy predictions with unprecedented computational efficiency, enabling the simulation of larger systems and longer timescales. In this study, we train a TorchMD-Net model \citep{tholke2022torchmd} using the MatDeepLearn framework \citep{fung2021benchmarking, jia2024derivative}. The training dataset, curated from the Materials Project \citep{chen2022universal}, comprises 187,687 crystal structures with associated energies, forces, and stresses. The model is trained for 400 epochs on a single Nvidia A100 80GB GPU. 

\subsection{Machine Learning Property Predictor}
Similar to machine learning force fields (MLFFs), machine learning property predictors (MLPPs) leverage advanced machine learning models trained on large datasets to make fast and accurate predictions for specific target properties. In this study, we train TorchMD-Net models to predict two separate properties: band gap and formation energy per atom. The datasets used are the \texttt{mp\_gap} and \texttt{mp\_form} datasets from the MatBench benchmark \citep{dunn2020benchmarking}, containing 106,113 and 132,752 structures from the Materials Project \citep{chen2022universal}, respectively. Each model is trained for 200 epochs on a single Nvidia A100 80GB GPU.

\subsection{Modification of Material}
Once LLMatDesign suggests a modification to achieve the user's target objective, the material is modified accordingly. Specifically, as illustrated in Fig. \ref{fig:overview}, there are four types of modifications: \texttt{exchange}, \texttt{substitute}, \texttt{remove}, and \texttt{add}. Each modification is applied directly to an \texttt{ase.Atoms} object representing the material. For example, given the modification \texttt{[`exchange', `Sr', `Ti']}, all Sr atoms in the material are replaced with Ti atoms and vice versa. After applying the modification, the structure undergoes relaxation using a machine learning force field (MLFF).

\subsection{Modification of Material}
The DFT calculations were performed using the Vienna Ab Initio Simulation Package (VASP)\citep{kresse1996efficient, kresse1996efficiency}. All calculations followed the same settings specified by the "MPRelaxSet" in the Pymatgen library\citep{ong2013python} used in Materials Project.

\section{Data Availability}
The authors declare that the data, materials and code supporting the results reported in this study are available upon the publication of this manuscript.

\section{Acknowledgements}
We thank Lingkai Kong and Rui Feng for helpful discussions.

This research used resources of the National Energy Research Scientific Computing Center (NERSC), a U.S. Department of Energy Office of Science User Facility located at Lawrence Berkeley National Laboratory, operated under Contract No. DE-AC02-05CH11231 using NERSC award BES-ERCAP0022842.




\clearpage
\bibliography{refs}

\begin{thebibliography}{10}
\expandafter\ifx\csname url\endcsname\relax
  \def\url#1{\texttt{#1}}\fi
\expandafter\ifx\csname urlprefix\endcsname\relax\def\urlprefix{URL }\fi
\providecommand{\bibinfo}[2]{#2}
\providecommand{\eprint}[2][]{\url{#2}}

\bibitem{davies2016computational}
\bibinfo{author}{Davies, D.~W.} \emph{et~al.}
\newblock \bibinfo{title}{Computational screening of all stoichiometric inorganic materials}.
\newblock \emph{\bibinfo{journal}{Chem}} \textbf{\bibinfo{volume}{1}}, \bibinfo{pages}{617--627} (\bibinfo{year}{2016}).

\bibitem{oganov2019structure}
\bibinfo{author}{Oganov, A.~R.}, \bibinfo{author}{Pickard, C.~J.}, \bibinfo{author}{Zhu, Q.} \& \bibinfo{author}{Needs, R.~J.}
\newblock \bibinfo{title}{Structure prediction drives materials discovery}.
\newblock \emph{\bibinfo{journal}{Nature Reviews Materials}} \textbf{\bibinfo{volume}{4}}, \bibinfo{pages}{331--348} (\bibinfo{year}{2019}).

\bibitem{liu2017materials}
\bibinfo{author}{Liu, Y.}, \bibinfo{author}{Zhao, T.}, \bibinfo{author}{Ju, W.} \& \bibinfo{author}{Shi, S.}
\newblock \bibinfo{title}{Materials discovery and design using machine learning}.
\newblock \emph{\bibinfo{journal}{Journal of Materiomics}} \textbf{\bibinfo{volume}{3}}, \bibinfo{pages}{159--177} (\bibinfo{year}{2017}).

\bibitem{hautier2012computer}
\bibinfo{author}{Hautier, G.}, \bibinfo{author}{Jain, A.} \& \bibinfo{author}{Ong, S.~P.}
\newblock \bibinfo{title}{From the computer to the laboratory: materials discovery and design using first-principles calculations}.
\newblock \emph{\bibinfo{journal}{Journal of Materials Science}} \textbf{\bibinfo{volume}{47}}, \bibinfo{pages}{7317--7340} (\bibinfo{year}{2012}).

\bibitem{pyzer2015high}
\bibinfo{author}{Pyzer-Knapp, E.~O.}, \bibinfo{author}{Suh, C.}, \bibinfo{author}{G{\'o}mez-Bombarelli, R.}, \bibinfo{author}{Aguilera-Iparraguirre, J.} \& \bibinfo{author}{Aspuru-Guzik, A.}
\newblock \bibinfo{title}{What is high-throughput virtual screening? a perspective from organic materials discovery}.
\newblock \emph{\bibinfo{journal}{Annual Review of Materials Research}} \textbf{\bibinfo{volume}{45}}, \bibinfo{pages}{195--216} (\bibinfo{year}{2015}).

\bibitem{chen2022universal}
\bibinfo{author}{Chen, C.} \& \bibinfo{author}{Ong, S.~P.}
\newblock \bibinfo{title}{A universal graph deep learning interatomic potential for the periodic table}.
\newblock \emph{\bibinfo{journal}{Nature Computational Science}} \textbf{\bibinfo{volume}{2}}, \bibinfo{pages}{718--728} (\bibinfo{year}{2022}).

\bibitem{merchant2023scaling}
\bibinfo{author}{Merchant, A.} \emph{et~al.}
\newblock \bibinfo{title}{Scaling deep learning for materials discovery}.
\newblock \emph{\bibinfo{journal}{Nature}} \textbf{\bibinfo{volume}{624}}, \bibinfo{pages}{80--85} (\bibinfo{year}{2023}).

\bibitem{hoffmann2019data}
\bibinfo{author}{Hoffmann, J.} \emph{et~al.}
\newblock \bibinfo{title}{Data-driven approach to encoding and decoding 3-d crystal structures}.
\newblock \emph{\bibinfo{journal}{arXiv preprint arXiv:1909.00949}}  (\bibinfo{year}{2019}).

\bibitem{court20203}
\bibinfo{author}{Court, C.~J.}, \bibinfo{author}{Yildirim, B.}, \bibinfo{author}{Jain, A.} \& \bibinfo{author}{Cole, J.~M.}
\newblock \bibinfo{title}{3-d inorganic crystal structure generation and property prediction via representation learning}.
\newblock \emph{\bibinfo{journal}{Journal of Chemical Information and Modeling}} \textbf{\bibinfo{volume}{60}}, \bibinfo{pages}{4518--4535} (\bibinfo{year}{2020}).

\bibitem{xie2021crystal}
\bibinfo{author}{Xie, T.}, \bibinfo{author}{Fu, X.}, \bibinfo{author}{Ganea, O.-E.}, \bibinfo{author}{Barzilay, R.} \& \bibinfo{author}{Jaakkola, T.}
\newblock \bibinfo{title}{Crystal diffusion variational autoencoder for periodic material generation}.
\newblock \emph{\bibinfo{journal}{arXiv preprint arXiv:2110.06197}}  (\bibinfo{year}{2021}).

\bibitem{long2021constrained}
\bibinfo{author}{Long, T.} \emph{et~al.}
\newblock \bibinfo{title}{Constrained crystals deep convolutional generative adversarial network for the inverse design of crystal structures}.
\newblock \emph{\bibinfo{journal}{npj Computational Materials}} \textbf{\bibinfo{volume}{7}}, \bibinfo{pages}{66} (\bibinfo{year}{2021}).

\bibitem{ren2022invertible}
\bibinfo{author}{Ren, Z.} \emph{et~al.}
\newblock \bibinfo{title}{An invertible crystallographic representation for general inverse design of inorganic crystals with targeted properties}.
\newblock \emph{\bibinfo{journal}{Matter}} \textbf{\bibinfo{volume}{5}}, \bibinfo{pages}{314--335} (\bibinfo{year}{2022}).

\bibitem{fung2022atomic}
\bibinfo{author}{Fung, V.} \emph{et~al.}
\newblock \bibinfo{title}{Atomic structure generation from reconstructing structural fingerprints}.
\newblock \emph{\bibinfo{journal}{Machine Learning: Science and Technology}} \textbf{\bibinfo{volume}{3}}, \bibinfo{pages}{045018} (\bibinfo{year}{2022}).

\bibitem{zeni2023mattergen}
\bibinfo{author}{Zeni, C.} \emph{et~al.}
\newblock \bibinfo{title}{Mattergen: a generative model for inorganic materials design}.
\newblock \emph{\bibinfo{journal}{arXiv preprint arXiv:2312.03687}}  (\bibinfo{year}{2023}).

\bibitem{wei2022chain}
\bibinfo{author}{Wei, J.} \emph{et~al.}
\newblock \bibinfo{title}{Chain-of-thought prompting elicits reasoning in large language models}.
\newblock \emph{\bibinfo{journal}{Advances in neural information processing systems}} \textbf{\bibinfo{volume}{35}}, \bibinfo{pages}{24824--24837} (\bibinfo{year}{2022}).

\bibitem{huang2022towards}
\bibinfo{author}{Huang, J.} \& \bibinfo{author}{Chang, K. C.-C.}
\newblock \bibinfo{title}{Towards reasoning in large language models: A survey}.
\newblock \emph{\bibinfo{journal}{arXiv preprint arXiv:2212.10403}}  (\bibinfo{year}{2022}).

\bibitem{li2022pre}
\bibinfo{author}{Li, S.} \emph{et~al.}
\newblock \bibinfo{title}{Pre-trained language models for interactive decision-making}.
\newblock \emph{\bibinfo{journal}{Advances in Neural Information Processing Systems}} \textbf{\bibinfo{volume}{35}}, \bibinfo{pages}{31199--31212} (\bibinfo{year}{2022}).

\bibitem{ahn2022can}
\bibinfo{author}{Ahn, M.} \emph{et~al.}
\newblock \bibinfo{title}{Do as i can, not as i say: Grounding language in robotic affordances}.
\newblock \emph{\bibinfo{journal}{arXiv preprint arXiv:2204.01691}}  (\bibinfo{year}{2022}).

\bibitem{huang2023voxposer}
\bibinfo{author}{Huang, W.} \emph{et~al.}
\newblock \bibinfo{title}{Voxposer: Composable 3d value maps for robotic manipulation with language models}.
\newblock \emph{\bibinfo{journal}{arXiv preprint arXiv:2307.05973}}  (\bibinfo{year}{2023}).

\bibitem{boiko2023autonomous}
\bibinfo{author}{Boiko, D.~A.}, \bibinfo{author}{MacKnight, R.}, \bibinfo{author}{Kline, B.} \& \bibinfo{author}{Gomes, G.}
\newblock \bibinfo{title}{Autonomous chemical research with large language models}.
\newblock \emph{\bibinfo{journal}{Nature}} \textbf{\bibinfo{volume}{624}}, \bibinfo{pages}{570--578} (\bibinfo{year}{2023}).

\bibitem{bran2023chemcrow}
\bibinfo{author}{Bran, A.~M.} \emph{et~al.}
\newblock \bibinfo{title}{Chemcrow: Augmenting large-language models with chemistry tools}.
\newblock \emph{\bibinfo{journal}{arXiv preprint arXiv:2304.05376}}  (\bibinfo{year}{2023}).

\bibitem{ai4science2023impact}
\bibinfo{author}{AI4Science, M.~R.} \& \bibinfo{author}{Quantum, M.~A.}
\newblock \bibinfo{title}{The impact of large language models on scientific discovery: a preliminary study using gpt-4}.
\newblock \emph{\bibinfo{journal}{arXiv preprint arXiv:2311.07361}}  (\bibinfo{year}{2023}).

\bibitem{mirza2024large}
\bibinfo{author}{Mirza, A.} \emph{et~al.}
\newblock \bibinfo{title}{Are large language models superhuman chemists?}
\newblock \emph{\bibinfo{journal}{arXiv preprint arXiv:2404.01475}}  (\bibinfo{year}{2024}).

\bibitem{fan2024openchemie}
\bibinfo{author}{Fan, V.} \emph{et~al.}
\newblock \bibinfo{title}{Openchemie: An information extraction toolkit for chemistry literature}.
\newblock \emph{\bibinfo{journal}{arXiv preprint arXiv:2404.01462}}  (\bibinfo{year}{2024}).

\bibitem{ai2024extracting}
\bibinfo{author}{Ai, Q.}, \bibinfo{author}{Meng, F.}, \bibinfo{author}{Shi, J.}, \bibinfo{author}{Pelkie, B.} \& \bibinfo{author}{Coley, C.~W.}
\newblock \bibinfo{title}{Extracting structured data from organic synthesis procedures using a fine-tuned large language model}.
\newblock \emph{\bibinfo{journal}{ChemRxiv preprint 10.26434/chemrxiv-2024-979fz}}  (\bibinfo{year}{2024}).

\bibitem{zhong2024benchmarking}
\bibinfo{author}{Zhong, Z.}, \bibinfo{author}{Zhou, K.} \& \bibinfo{author}{Mottin, D.}
\newblock \bibinfo{title}{Benchmarking large language models for molecule prediction tasks}.
\newblock \emph{\bibinfo{journal}{arXiv preprint arXiv:2403.05075}}  (\bibinfo{year}{2024}).

\bibitem{xie2024fine}
\bibinfo{author}{Xie, Z.} \emph{et~al.}
\newblock \bibinfo{title}{Fine-tuning gpt-3 for machine learning electronic and functional properties of organic molecules}.
\newblock \emph{\bibinfo{journal}{Chemical science}} \textbf{\bibinfo{volume}{15}}, \bibinfo{pages}{500--510} (\bibinfo{year}{2024}).

\bibitem{jablonka2024leveraging}
\bibinfo{author}{Jablonka, K.~M.}, \bibinfo{author}{Schwaller, P.}, \bibinfo{author}{Ortega-Guerrero, A.} \& \bibinfo{author}{Smit, B.}
\newblock \bibinfo{title}{Leveraging large language models for predictive chemistry}.
\newblock \emph{\bibinfo{journal}{Nature Machine Intelligence}} \bibinfo{pages}{1--9} (\bibinfo{year}{2024}).

\bibitem{ock2023catalyst}
\bibinfo{author}{Ock, J.}, \bibinfo{author}{Guntuboina, C.} \& \bibinfo{author}{Barati~Farimani, A.}
\newblock \bibinfo{title}{Catalyst energy prediction with catberta: Unveiling feature exploration strategies through large language models}.
\newblock \emph{\bibinfo{journal}{ACS Catalysis}} \textbf{\bibinfo{volume}{13}}, \bibinfo{pages}{16032--16044} (\bibinfo{year}{2023}).

\bibitem{flam2023language}
\bibinfo{author}{Flam-Shepherd, D.} \& \bibinfo{author}{Aspuru-Guzik, A.}
\newblock \bibinfo{title}{Language models can generate molecules, materials, and protein binding sites directly in three dimensions as xyz, cif, and pdb files}.
\newblock \emph{\bibinfo{journal}{arXiv preprint arXiv:2305.05708}}  (\bibinfo{year}{2023}).

\bibitem{antunes2023crystal}
\bibinfo{author}{Antunes, L.~M.}, \bibinfo{author}{Butler, K.~T.} \& \bibinfo{author}{Grau-Crespo, R.}
\newblock \bibinfo{title}{Crystal structure generation with autoregressive large language modeling}.
\newblock \emph{\bibinfo{journal}{arXiv preprint arXiv:2307.04340}}  (\bibinfo{year}{2023}).

\bibitem{gruver2024fine}
\bibinfo{author}{Gruver, N.} \emph{et~al.}
\newblock \bibinfo{title}{Fine-tuned language models generate stable inorganic materials as text}.
\newblock \emph{\bibinfo{journal}{arXiv preprint arXiv:2402.04379}}  (\bibinfo{year}{2024}).

\bibitem{jain2013commentary}
\bibinfo{author}{Jain, A.} \emph{et~al.}
\newblock \bibinfo{title}{Commentary: The materials project: A materials genome approach to accelerating materials innovation}.
\newblock \emph{\bibinfo{journal}{APL materials}} \textbf{\bibinfo{volume}{1}} (\bibinfo{year}{2013}).

\bibitem{achiam2023gpt}
\bibinfo{author}{Achiam, J.} \emph{et~al.}
\newblock \bibinfo{title}{Gpt-4 technical report}.
\newblock \emph{\bibinfo{journal}{arXiv preprint arXiv:2303.08774}}  (\bibinfo{year}{2023}).

\bibitem{team2023gemini}
\bibinfo{author}{Team, G.} \emph{et~al.}
\newblock \bibinfo{title}{Gemini: a family of highly capable multimodal models}.
\newblock \emph{\bibinfo{journal}{arXiv preprint arXiv:2312.11805}}  (\bibinfo{year}{2023}).

\bibitem{sutherland2020solar}
\bibinfo{author}{Sutherland, B.~R.}
\newblock \bibinfo{title}{Solar materials find their band gap}.
\newblock \emph{\bibinfo{journal}{Joule}} \textbf{\bibinfo{volume}{4}}, \bibinfo{pages}{984--985} (\bibinfo{year}{2020}).

\bibitem{vaswani2017attention}
\bibinfo{author}{Vaswani, A.} \emph{et~al.}
\newblock \bibinfo{title}{Attention is all you need}.
\newblock \emph{\bibinfo{journal}{Advances in neural information processing systems}} \textbf{\bibinfo{volume}{30}} (\bibinfo{year}{2017}).

\bibitem{ko2023recent}
\bibinfo{author}{Ko, T.~W.} \& \bibinfo{author}{Ong, S.~P.}
\newblock \bibinfo{title}{Recent advances and outstanding challenges for machine learning interatomic potentials}.
\newblock \emph{\bibinfo{journal}{Nature Computational Science}} \textbf{\bibinfo{volume}{3}}, \bibinfo{pages}{998--1000} (\bibinfo{year}{2023}).

\bibitem{tholke2022torchmd}
\bibinfo{author}{Th{\"o}lke, P.} \& \bibinfo{author}{De~Fabritiis, G.}
\newblock \bibinfo{title}{Torchmd-net: Equivariant transformers for neural network based molecular potentials}.
\newblock \emph{\bibinfo{journal}{arXiv preprint arXiv:2202.02541}}  (\bibinfo{year}{2022}).

\bibitem{fung2021benchmarking}
\bibinfo{author}{Fung, V.}, \bibinfo{author}{Zhang, J.}, \bibinfo{author}{Juarez, E.} \& \bibinfo{author}{Sumpter, B.~G.}
\newblock \bibinfo{title}{Benchmarking graph neural networks for materials chemistry}.
\newblock \emph{\bibinfo{journal}{npj Computational Materials}} \textbf{\bibinfo{volume}{7}}, \bibinfo{pages}{84} (\bibinfo{year}{2021}).

\bibitem{jia2024derivative}
\bibinfo{author}{Jia, S.} \emph{et~al.}
\newblock \bibinfo{title}{Derivative-based pre-training of graph neural networks for materials property predictions}.
\newblock \emph{\bibinfo{journal}{Digital Discovery}} \textbf{\bibinfo{volume}{3}}, \bibinfo{pages}{586--593} (\bibinfo{year}{2024}).

\bibitem{dunn2020benchmarking}
\bibinfo{author}{Dunn, A.}, \bibinfo{author}{Wang, Q.}, \bibinfo{author}{Ganose, A.}, \bibinfo{author}{Dopp, D.} \& \bibinfo{author}{Jain, A.}
\newblock \bibinfo{title}{Benchmarking materials property prediction methods: the matbench test set and automatminer reference algorithm}.
\newblock \emph{\bibinfo{journal}{npj Computational Materials}} \textbf{\bibinfo{volume}{6}}, \bibinfo{pages}{138} (\bibinfo{year}{2020}).

\bibitem{kresse1996efficient}
\bibinfo{author}{Kresse, G.} \& \bibinfo{author}{Furthm{\"u}ller, J.}
\newblock \bibinfo{title}{Efficient iterative schemes for ab initio total-energy calculations using a plane-wave basis set}.
\newblock \emph{\bibinfo{journal}{Physical review B}} \textbf{\bibinfo{volume}{54}}, \bibinfo{pages}{11169} (\bibinfo{year}{1996}).

\bibitem{kresse1996efficiency}
\bibinfo{author}{Kresse, G.} \& \bibinfo{author}{Furthm{\"u}ller, J.}
\newblock \bibinfo{title}{Efficiency of ab-initio total energy calculations for metals and semiconductors using a plane-wave basis set}.
\newblock \emph{\bibinfo{journal}{Computational materials science}} \textbf{\bibinfo{volume}{6}}, \bibinfo{pages}{15--50} (\bibinfo{year}{1996}).

\bibitem{ong2013python}
\bibinfo{author}{Ong, S.~P.} \emph{et~al.}
\newblock \bibinfo{title}{Python materials genomics (pymatgen): A robust, open-source python library for materials analysis}.
\newblock \emph{\bibinfo{journal}{Computational Materials Science}} \textbf{\bibinfo{volume}{68}}, \bibinfo{pages}{314--319} (\bibinfo{year}{2013}).

\end{thebibliography}
\bibliographystyle{naturemag}
\clearpage

\appendix
\begin{titlepage}
  \centering
  \LARGE \textsc{Supplementary Information} \par
  \let\endtitlepage\relax
\end{titlepage}

\counterwithin{figure}{section}
\counterwithin{table}{section}

\section{Prompt Templates for LLMatDesign}\label{sec:appendix-prompts}
A slightly different prompt template to Fig. \ref{fig:basic_prompt} is designed for Gemini-1.0-pro due to its inconsistency in generating standardized output. 

\noindent\begin{minipage}{\textwidth}
\begin{tcolorbox}[
    colback=white, colframe=black, coltitle=black, colbacktitle={rgb,255:red, 254; green, 229; blue, 178},
    title=LLMatDesign Prompt Template (Gemini-1.0-pro), 
    fontupper=\footnotesize, 
    fontlower=\footnotesize
]
    I have a material and its {\color{red} \texttt{<property>}}. {\color{red}\texttt{<definition of property>}}.\newline
    
    ({\color{red}\texttt{<chemical composition>}}, {\color{red}\texttt{<property value>}})\newline

    You will be given a starting material to be modified. Try to achieve {\color{red}\texttt{<objective>}}. Make an informed choice of modification based on the given material and past modifications and property values obtained after those modifications. Output a list for the suggested modification, and a string of the reason why you think it is a good modification to take to achieve {\color{red}\texttt{<objective>}}. Make sure the modification is physically meaningful.\newline

    Material to be modified: {\color{red}\texttt{<chemical composition>}}
    
    Current property value: {\color{red}\texttt{<property value>}}\newline

    {\color{blue}\texttt{<modification history>}}\newline 

    Available modifications:    
    \begin{enumerate}[noitemsep, leftmargin=*]
        \item exchange: exchange two elements in the material
        \item substitute: substitute one element in the material with another 
        \item remove: remove an element from the material 
        \item add: add an element to the material
    \end{enumerate}

    Example output format:
    \begin{enumerate}[noitemsep, leftmargin=*]
        \item \texttt{["exchange", "O", "N"], "some reason here"}
        \item \texttt{["substitute", "Ti", "Fe"], "some reason here"}
        \item \texttt{["add", "O"], "some reason here"}
        \item \texttt{["remove", "O"], "some reason here"}
    \end{enumerate}
\end{tcolorbox}
\captionof{figure}{Prompt template for LLMatDesign with Gemini-1.0-pro. Text placeholders in red angular brackets are specific to the task given to LLMatDesign. Text placeholders in blue angular brackets are optional and can be omitted if not needed.}\label{fig:basic_prompt_gemini}
\end{minipage}
\clearpage

\noindent\begin{minipage}[t]{\textwidth}
\begin{tcolorbox}[
    colback=white, colframe=black, coltitle=black, colbacktitle={rgb,255:red, 254; green, 229; blue, 178},
    title=GPT-4o Refined Prompt Template for LLMatDesign, 
    fontupper=\footnotesize, 
    fontlower=\footnotesize
]
    I have a material with a known {\color{red} \texttt{<property>}}. {\color{red}\texttt{<definition of property>}}.\newline

    Material information:
    \begin{itemize}[noitemsep, leftmargin=*]
        \item Chemical formula: {\color{red}\texttt{<chemical composition>}}
        \item {\color{red} \texttt{<property>}}: {\color{red}\texttt{<property value>}}
    \end{itemize}

    Objective:

    Propose a modification to this material to achieve {\color{red}\texttt{<objective>}}. You can choose one of the following modification types:

    \begin{enumerate}[noitemsep, leftmargin=*]
        \item exchange: exchange two elements in the material
        \item substitute: substitute one element in the material with another 
        \item remove: remove an element from the material 
        \item add: add an element to the material
    \end{enumerate}
    
    Your response should be a Python dictionary in the following format:\newline

    \texttt{```}
    
    \{Hypothesis: \textdollar HYPOTHESIS, Modification: [\textdollar TYPE, \textdollar ELEMENT\_1, \textdollar ELEMENT\_2]\}.

    \texttt{```}\newline
    
    Requirements:
    \begin{enumerate}[noitemsep, leftmargin=*]
        \item \textdollar HYPOTHESIS: Provide a detailed analysis and rationale for your proposed modification.
        \item \textdollar TYPE:Specify the type of modification (``exchange", ``substitute", ``remove", ``add"). 
        \item \textdollar Identify the element(s) involved in the modification. For "exchange" and "substitute", include two elements (\textdollar ELEMENT\_1 and \textdollar ELEMENT\_2). For ``remove" and ``add", include one element (\textdollar ELEMENT\_1).
    \end{enumerate}

    {\color{blue}\texttt{<modification history>}}\newline

    Take a deep breath and work on this problem step-by-step. Your thoughtful and detailed analysis is highly appreciated.
\end{tcolorbox}
\captionof{figure}{GPT-4o refined prompt template for LLMatDesign. Text placeholders in red angular brackets are specific to the task given to LLMatDesign. Text placeholders in blue angular brackets are optional and can be omitted if not needed.}\label{fig:gpt_4o_refined_prompt}
\end{minipage}

\noindent\begin{minipage}[t]{\textwidth}
\begin{tcolorbox}[
    colback=white, colframe=black, coltitle=black, colbacktitle={rgb,255:red, 254; green, 229; blue, 178},
    title=Persona Prompt Template for LLMatDesign, 
    fontupper=\footnotesize, 
    fontlower=\footnotesize
]
    You are a materials design expert working on the development of new materials with specific properties. You will be given a composition (chemical formula) and its corresponding {\color{red} \texttt{<property>}}. You will be asked to propose a modification to the material to achieve a target {\color{red} \texttt{<property>}}.\newline

    Material information:
    \begin{itemize}[noitemsep, leftmargin=*]
        \item Chemical formula: {\color{red}\texttt{<chemical composition>}}
        \item {\color{red} \texttt{<property>}}: {\color{red}\texttt{<property value>}}
    \end{itemize}

    Objective:

    Propose a modification to this material to achieve {\color{red}\texttt{<objective>}}. You can choose one of the following modification types:

    \begin{enumerate}[noitemsep, leftmargin=*]
        \item exchange: exchange two elements in the material
        \item substitute: substitute one element in the material with another 
        \item remove: remove an element from the material 
        \item add: add an element to the material
    \end{enumerate}
    
    Your response should be a Python dictionary in the following format:\newline

    \texttt{```}
    
    \{Hypothesis: \textdollar HYPOTHESIS, Modification: [\textdollar TYPE, \textdollar ELEMENT\_1, \textdollar ELEMENT\_2]\}.

    \texttt{```}\newline
    
    Requirements:
    \begin{enumerate}[noitemsep, leftmargin=*]
        \item \textdollar HYPOTHESIS: Provide a detailed analysis and rationale for your proposed modification.
        \item \textdollar TYPE:Specify the type of modification (``exchange", ``substitute", ``remove", ``add"). 
        \item \textdollar Identify the element(s) involved in the modification. For "exchange" and "substitute", include two elements (\textdollar ELEMENT\_1 and \textdollar ELEMENT\_2). For ``remove" and ``add", include one element (\textdollar ELEMENT\_1).
    \end{enumerate}

    {\color{blue}\texttt{<modification history>}}\newline

    Take a deep breath and work on this problem step-by-step. Your thoughtful and detailed analysis is highly appreciated.
\end{tcolorbox}
\captionof{figure}{Prompt template with materials design expert persona for LLMatDesign. Text placeholders in red angular brackets are specific to the task given to LLMatDesign. Text placeholders in blue angular brackets are optional and can be omitted if not needed.}\label{fig:persona_prompt}
\end{minipage}
\clearpage

\section{Convergence Plots}\label{sec:appendix-individual-plots}
\begin{figure}[h]
    \centering
    \includegraphics[width=\textwidth]{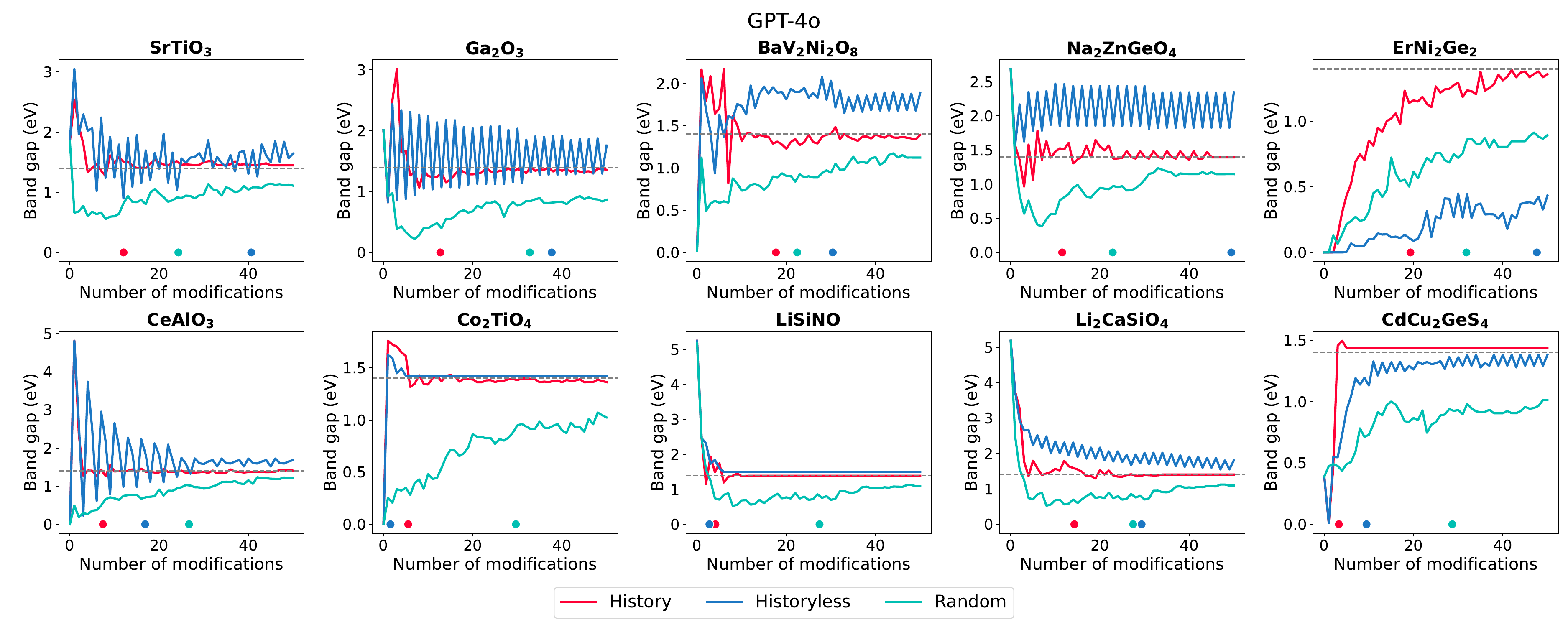}
    \caption{Average band gaps over 50 modifications for all 10 starting materials using GPT-4o. The grey horizontal line indicates the target band gap of 1.4 eV. The colored dots on the x-axis indicate the average number of modifications taken for each method to reach the target.}
    \label{fig:gpt4o-individual-bandgap}
\end{figure}

\begin{figure}[h]
    \centering
    \includegraphics[width=\textwidth]{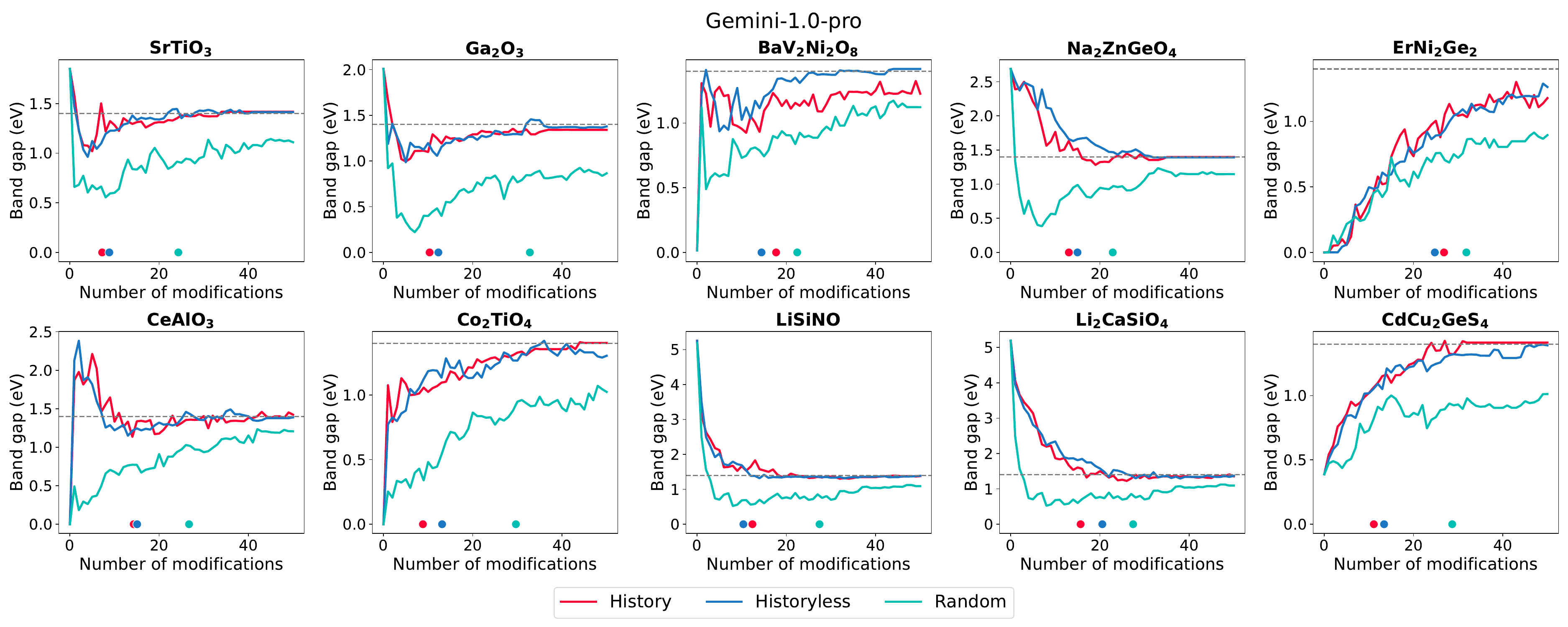}
    \caption{Average band gaps over 50 modifications for all 10 starting materials using Gemini-1.0-pro. The grey horizontal line indicates the target band gap of 1.4 eV. The colored dots on the x-axis indicate the average number of modifications taken for each method to reach the target.}
    \label{fig:gemini-individual-bandgap}
\end{figure}

\begin{figure}[H]
    \centering
    \includegraphics[width=\textwidth]{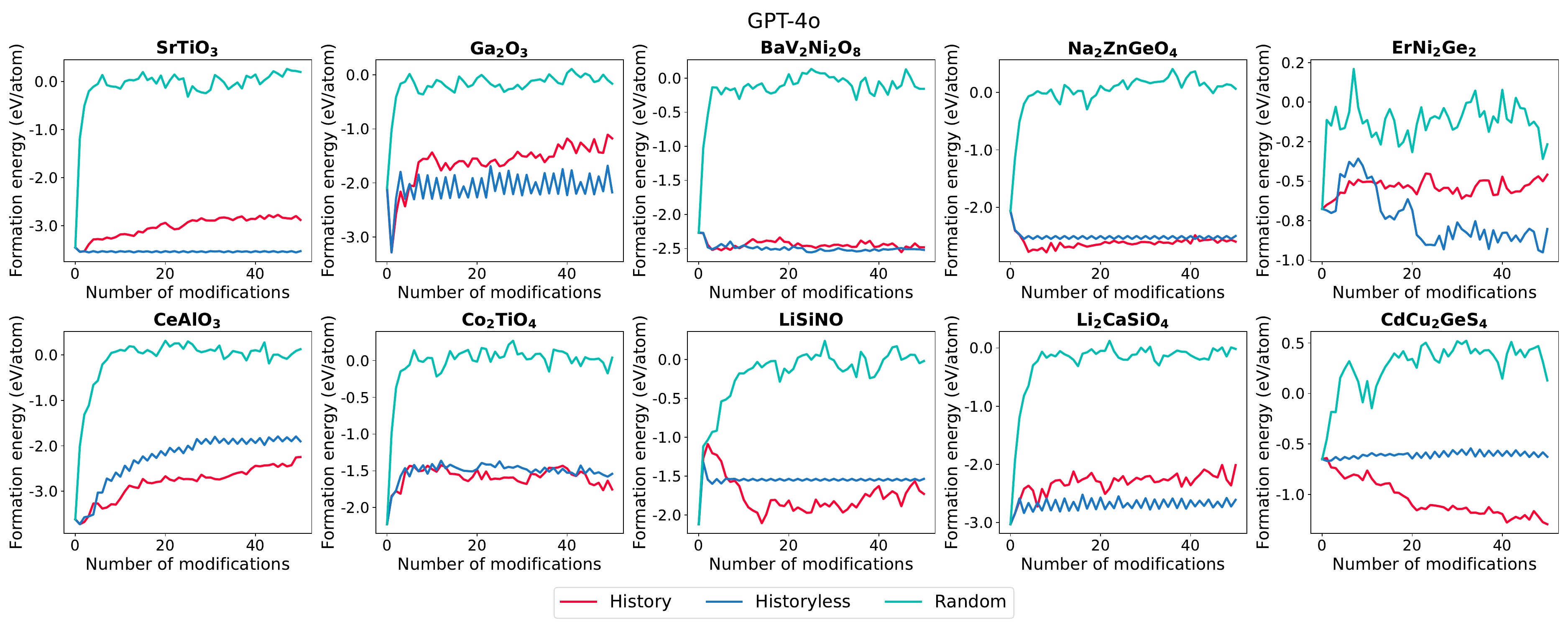}
    \caption{Average formation energies over 50 modifications for all 10 starting materials using GPT-4o. The goal is to achieve the lowest possible formation energy per atom.}
    \label{fig:gpt4o-individual-fe}
\end{figure}

\begin{figure}[H]
    \centering
    \includegraphics[width=\textwidth]{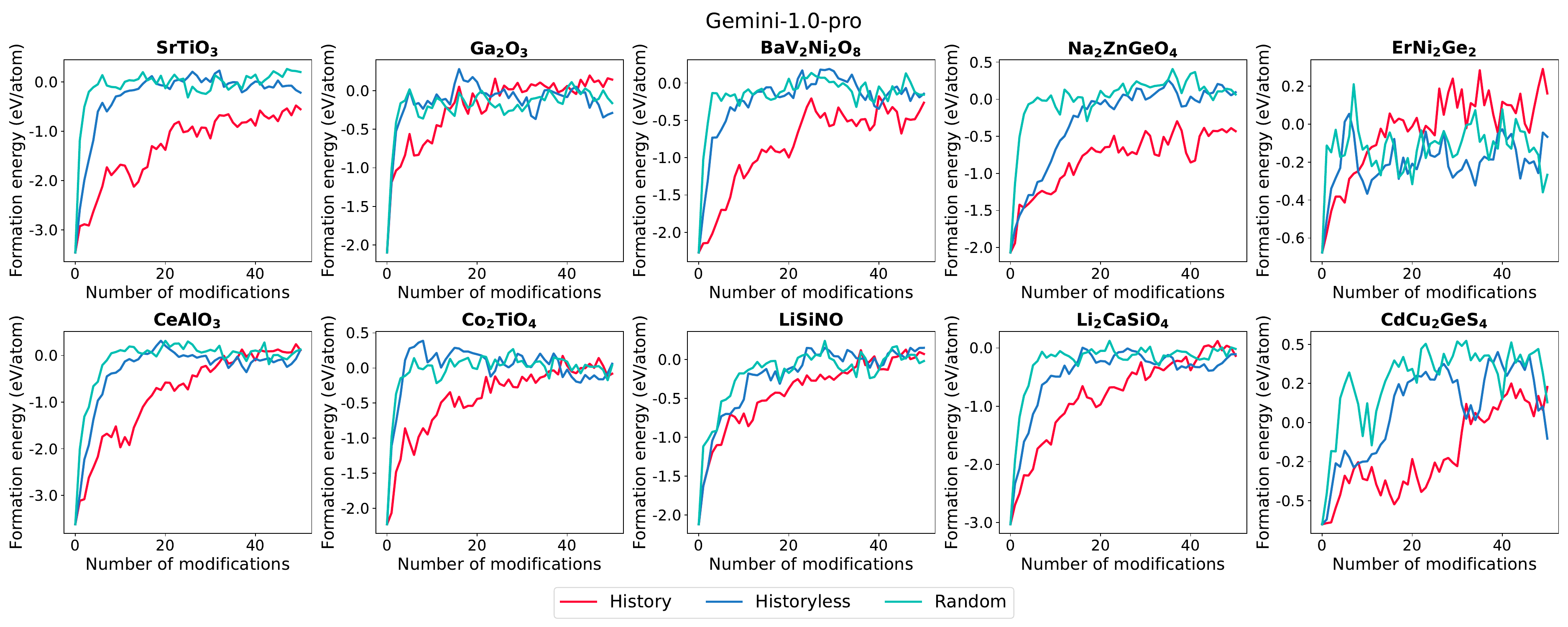}
    \caption{Average formation energies over 50 modifications for all 10 starting materials using Gemini-1.0-pro. The goal is to achieve the lowest possible formation energy per atom.}
    \label{fig:gemini-individual-fe}
\end{figure}

\clearpage

\section{Heatmaps}\label{sec:appendix-heatmaps}
\begin{figure}[h]
    \centering
    \begin{minipage}[b]{0.45\textwidth}
        \centering
        \includegraphics[width=\textwidth]{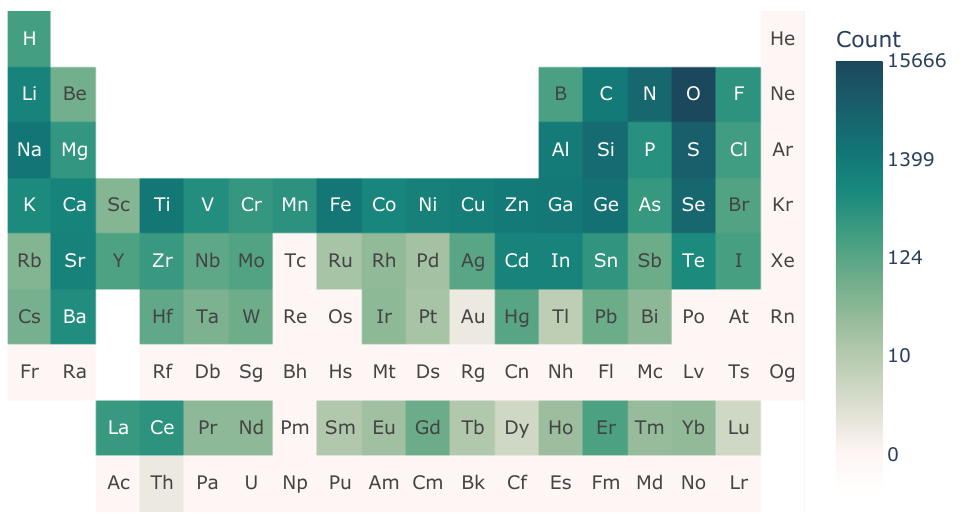}
        \captionsetup{font=small}
        \caption*{BG: Gemini-1.0-pro without history}
        \label{fig:periodic-BG-historyless-gemini-1.0-pro}
    \end{minipage}
    \hfill
    \begin{minipage}[b]{0.45\textwidth}
        \centering
        \includegraphics[width=\textwidth]{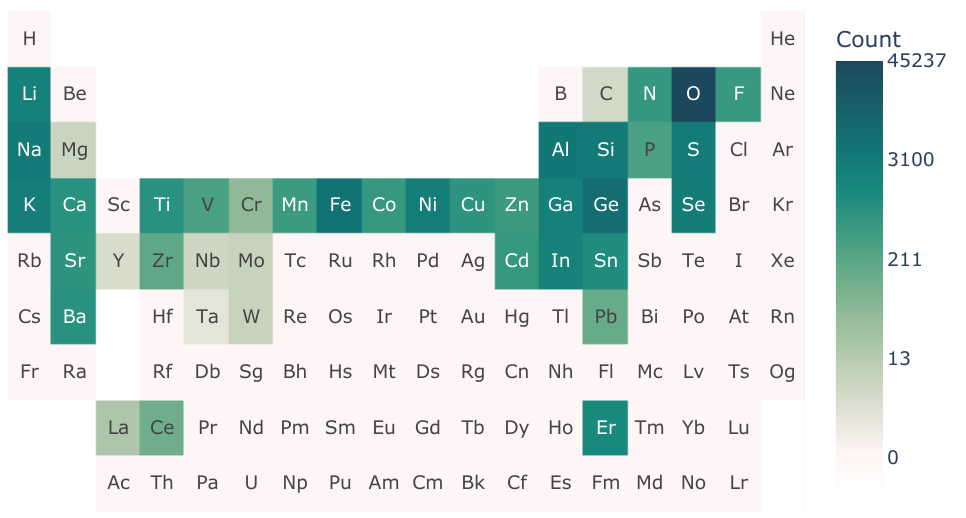}
        \captionsetup{font=small}
        \caption*{BG: GPT-4o without history}
        \label{fig:periodic-BG-historyless-GPT-4o}
    \end{minipage}

    \begin{minipage}[b]{0.45\textwidth}
        \centering
        \includegraphics[width=\textwidth]{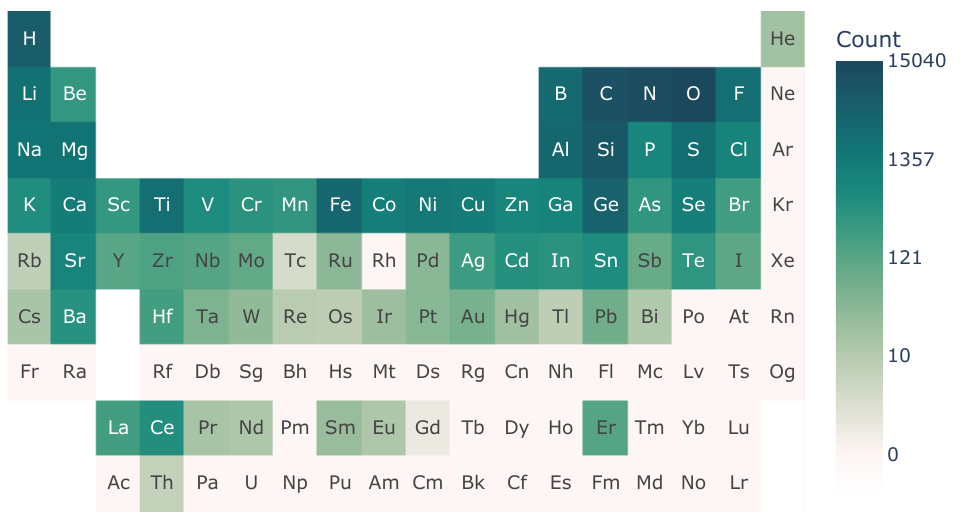}
        \captionsetup{font=small}
        \caption*{FE: Gemini-1.0-pro without history}
        \label{fig:periodic-FE-historyless-gemini-1.0-pro}
    \end{minipage}
    \hfill
    \begin{minipage}[b]{0.45\textwidth}
        \centering
        \includegraphics[width=\textwidth]{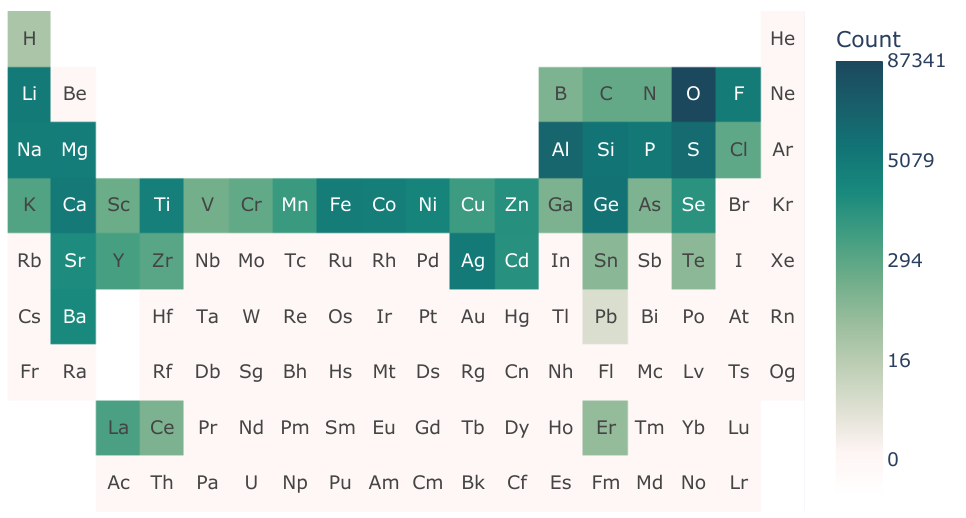}
        \captionsetup{font=small}
        \caption*{FE: GPT-4o without history}
        \label{fig:periodic-FE-historyless-GPT-4o}
    \end{minipage}
    
    \caption{Heatmaps of element frequencies in band gap (BG) and formation energy (FE) tasks for Gemini-1.0-pro and GPT-4o without history. The periodic table is color-coded to indicate the frequency of each element's occurrence in all modified materials (both intermediate and final) across all runs and starting materials. Darker colors represent higher frequencies, while lighter colors denote lower frequencies or absence. The visualization employs log-scaling to effectively highlight the distribution and prevalence of elements.}
    \label{fig:periodic_tables_heatmaps_historyless}
\end{figure}

\clearpage

\section{Visualization of Selected Structures}
\begin{figure}[h]
    \centering
    \begin{overpic}[width=\textwidth]{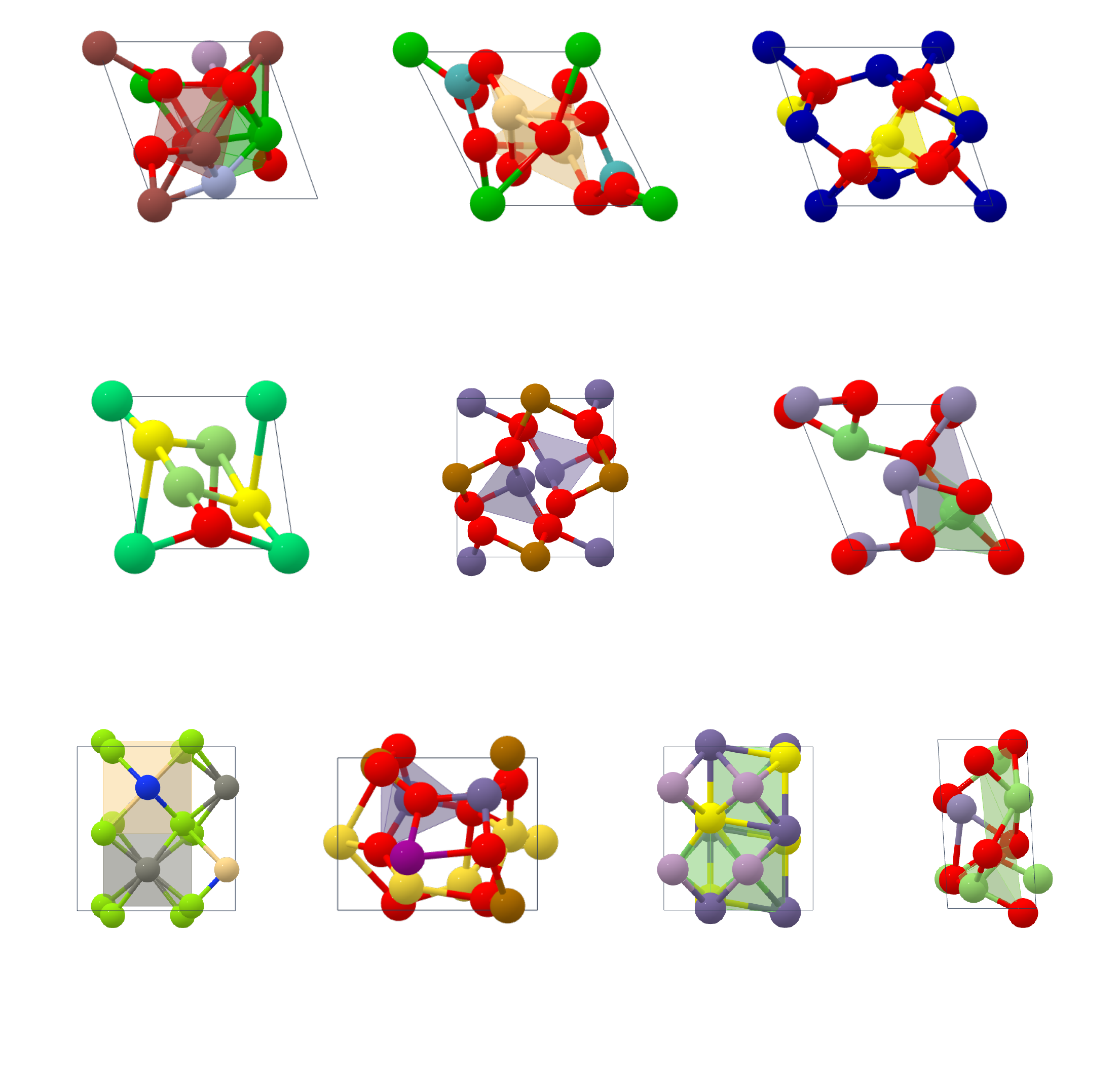}
        \put(10, 73){\textcolor{red}{SrTiO$_3$}}
        \put(10, 70){{Ba$_2$Tl$_2$PNO$_6$}}
        \put(10, 67){1.51 eV}

        \put(40, 73){\textcolor{red}{BaV$_2$Ni$_2$O$_8$}}
        \put(40, 70){BaCd$_2$(MoO$_5$)$_2$}
        \put(40, 67){1.30 eV}

        \put(72, 73){\textcolor{red}{Co$_2$TiO$_4$}}
        \put(72, 70){Co$_2$SO$_4$}
        \put(72, 67){1.42 eV}

        \put(10, 41){\textcolor{red}{ErNi$_2$Ge$_2$}}
        \put(10, 38){ErGa$_2$S$_2$O}
        \put(10, 35){1.39 eV}

        \put(40, 41){\textcolor{red}{CeAlO$_3$}}
        \put(40, 38){FeGeO$_3$}
        \put(40, 35){1.49 eV}

        \put(72, 41){\textcolor{red}{Li$_2$CaSiO$_4$}}
        \put(72, 38){LiSnO$_2$}
        \put(72, 35){1.29 eV}

        \put(7, 10){\textcolor{red}{CdCu$2$GeS$4$}}
        \put(7, 7){Zn$_2$CdSiSe$_4$}
        \put(7, 4){1.42 eV}

        \put(30, 10){\textcolor{red}{Na$_2$ZnGeO$_4$}}
        \put(30, 7){Na$_4$MnFe$_2$(GeO$_4$)$_2$}
        \put(30, 4){1.41 eV}

        \put(59, 10){\textcolor{red}{LiSiNO}}
        \put(59, 7){LiGePS}
        \put(59, 4){1.30 eV}

        \put(84, 10){\textcolor{red}{Ga$_2$O$_3$}}
        \put(84, 7){Ga$_4$SnO$_6$}
        \put(84, 4){1.33 eV}

    \end{overpic}
    \caption{Visualization of the final structures obtained by LLMatDesign for the band gap task. These structures are obtained from the first run of all 10 starting materials. The chemical formulae in red represent the starting materials, followed by the formulae of the final structures and their corresponding band gaps. GPT-4o with history is utilized as the LLM engine.}
    \label{fig:band_vis}
\end{figure}

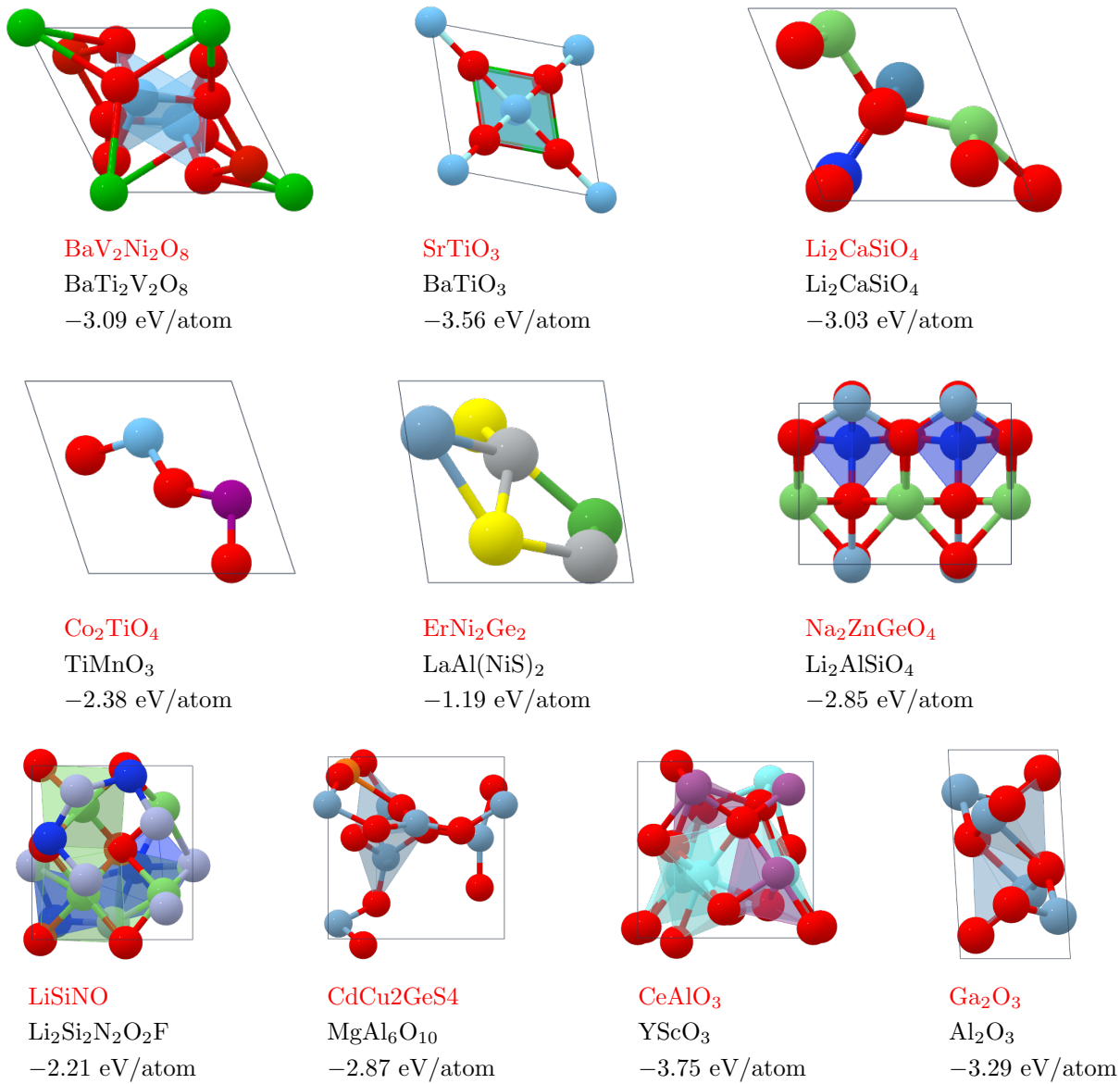
\begin{figure}[h]
    \centering
    \begin{overpic}[width=\textwidth]{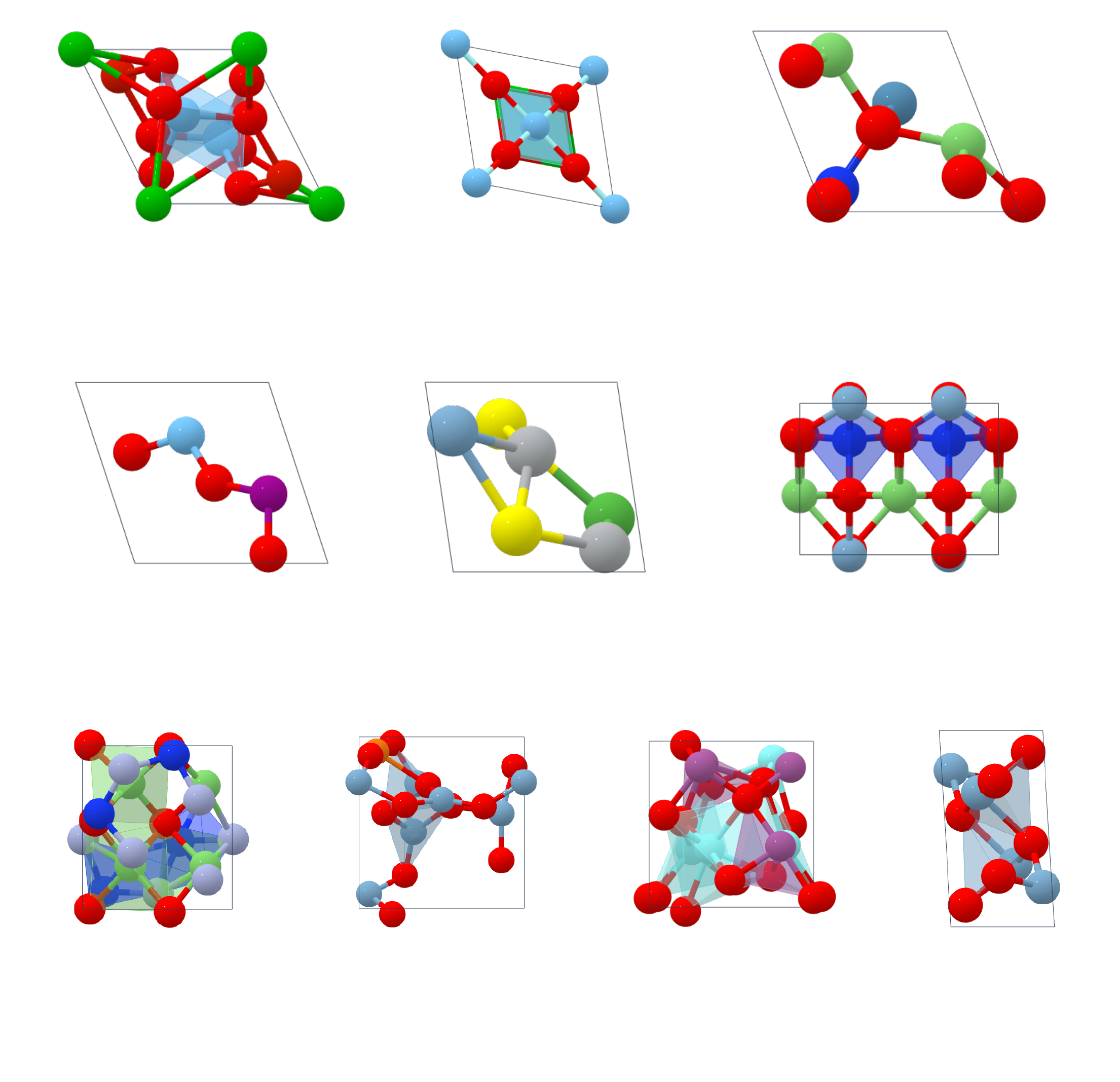}
        \put(10, 73){\textcolor{red}{BaV$_2$Ni$_2$O$_8$}}
        \put(10, 70){{BaTi$_2$V$_2$O$_8$}}
        \put(10, 67){$-3.09$ eV/atom}

        \put(40, 73){\textcolor{red}{SrTiO$_3$}}
        \put(40, 70){BaTiO$_3$}
        \put(40, 67){$-3.56$ eV/atom}

        \put(72, 73){\textcolor{red}{Li$_2$CaSiO$_4$}}
        \put(72, 70){Li$_2$CaSiO$_4$}
        \put(72, 67){$-3.03$ eV/atom}

        \put(10, 41){\textcolor{red}{Co$_2$TiO$_4$}}
        \put(10, 38){TiMnO$_3$}
        \put(10, 35){$-2.38$ eV/atom}

        \put(40, 41){\textcolor{red}{ErNi$_2$Ge$_2$}}
        \put(40, 38){LaAl(NiS)$_2$}
        \put(40, 35){$-1.19$ eV/atom}

        \put(72, 41){\textcolor{red}{Na$_2$ZnGeO$_4$}}
        \put(72, 38){Li$_2$AlSiO$_4$}
        \put(72, 35){$-2.85$ eV/atom}

        \put(7, 10){\textcolor{red}{LiSiNO}}
        \put(7, 7){Li$_2$Si$_2$N$_2$O$_2$F}
        \put(7, 4){$-2.21$ eV/atom}

        \put(32, 10){\textcolor{red}{CdCu$2$GeS$4$}}
        \put(32, 7){MgAl$_6$O$_{10}$}
        \put(32, 4){$-2.87$ eV/atom}

        \put(58, 10){\textcolor{red}{CeAlO$_3$}}
        \put(58, 7){YScO$_3$}
        \put(58, 4){$-3.75$ eV/atom}

        \put(84, 10){\textcolor{red}{Ga$_2$O$_3$}}
        \put(84, 7){Al$_2$O$_3$}
        \put(84, 4){$-3.29$ eV/atom}

    \end{overpic}
    \caption{Visualization of the final structures obtained by LLMatDesign for the formation energy task. These structures represent the ones with the minimum formation energy per atom from the first run of all 10 starting materials. The chemical formulae in red represent the starting materials, followed by the formulae of the structures with the lowest formation energies and their corresponding formation energy per atom values. GPT-4o with history is utilized as the LLM engine.}
    \label{fig:formation_vis}
\end{figure}
\clearpage

\section{DFT Calculations}
\begin{table}[ht]
\caption{ DFT results for lowest-energy structures obtained from the formation energy task, averaged across all starting materials and runs.}
\label{tab:dft-results-formation-energies}
\footnotesize
\begin{center}
\renewcommand\cellalign{c}
\setcellgapes{3pt}\makegapedcells
\begin{adjustbox}{max width=\textwidth}
\begin{tabular}{lcc}
\hline
                                & GPT-4o with history  & Random \\ \hline
Formation energy per atom (eV/atom)             & $-2.31$ &   $-1.51$                            \\
Job success rate (\%)                &  73.3        & 40.0 \\ \hline
\end{tabular}
\end{adjustbox}
\end{center}
\end{table}

\end{document}